# Near-equilibrium isotope fractionation during planetesimal evaporation

E. D. YOUNG*[1], A. SHAHAR[2], F. NIMMO[3], H. E. SCHLICHTING[1], E. A. SCHAUBLE[1], H. TANG[1], J. LABIDI[1]

[1]Department of Earth, Planetary, and Space Sciences, UCLA, Los Angeles, CA, USA, [*corresp: eyoung@epss.ucla.edu], [2]Geophysical Laboratory, Carnegie Institution for Science, Washington DC, USA, [3]Department of Earth and Planetary Sciences, University of California Santa Cruz, Santa Cruz CA, USA.

## Abstract

Silicon and Mg in differentiated rocky bodies exhibit heavy isotope enrichments that have been attributed to evaporation of partially or entirely molten planetesimals. We evaluate the mechanisms of planetesimal evaporation in the early solar system and the conditions that controled attendant isotope fractionations.

Energy balance at the surface of a body accreted within ~1 Myr of CAI formation and heated from within by $^{26}$Al decay results in internal temperatures exceeding the silicate solidus, producing a transient magma ocean with a thin surface boundary layer of order < 1 meter that would be subject to foundering. Bodies that are massive enough to form magma oceans by radioisotope decay ($\geq 0.1\%\ M_{\oplus}$) can retain hot rock vapor even in the absence of ambient nebular gas. We find that a steady-state rock vapor forms within minutes to hours and results from a balance between rates of magma evaporation and atmospheric escape. Vapor pressure buildup adjacent to the surfaces of the evaporating magmas would have inevitably led to an approach to equilibrium isotope partitioning between the vapor phase and the silicate melt. Numerical simulations of this near-equilibrium evaporation process for a body with a radius of ~ 700 km yield a steady-state far-field vapor pressure of $10^{-8}$ bar and a vapor pressure at the surface of $10^{-4}$ bar, corresponding to 95% saturation. Approaches to equilibrium isotope fractionation between vapor and melt should have been the norm during planet formation due to the formation of steady-state rock vapor atmospheres and/or the presence of protostellar gas.

We model the Si and Mg isotopic composition of bulk Earth as a consequence of accretion of planetesimals that evaporated subject to the conditions described above. The results show that the best fit to bulk Earth is for a carbonaceous chondrite-like source material with about 12% loss of Mg and 15% loss of Si resulting from near-equilibrium evaporation into the solar protostellar disk of $H_2$ on timescales of $10^4$ to $10^5$ years.

*Full contact details for corresponding author: Department of Earth, Planetary, and Space Sciences, UCLA, 595 Charles E. Young Drive East, Geology Building, Los Angeles, CA 90095; tel: 310 267-4930; Email: eyoung@epss.ucla.edu.

## 1. Introduction

Evaporation of magma exposed to space may have played a role in determining the chemical and isotopic compositions of rocky planets and their antecedents in the early solar system (e.g., Mittlefehldt 1987; Ringwood 1992; Halliday and Porcelli 2001). Identifying reliable signatures of evaporation of melts into space would therefore elucidate a potentially important process attending planet formation in general. Advances in our ability to measure stable isotope ratios of light, rock-forming elements, including those for Zn, K, Fe, Mg, and Si, among others, has resulted in new insights into planet formation in general. An emerging hypothesis is that evaporation of molten or partially molten rocky planetesimals, planetary embryos, and/or proto-planets[1] caused losses of moderately volatile elements (e.g., Zn and K) and "common" or moderately refractory elements (e.g., Mg and Si, with condensation temperatures > ~1300 K and < ~1580 K, Lodders 2003) during planet formation. The hypothesis is that Earth, the Moon, and other rocky bodies may have inherited chemical and isotopic signatures of evaporation from their source materials, or, in some cases, acquired them as the result of giant impacts (Schlichting and Mukhopadhyay 2018).

The primary evidence for evaporation during planet formation is the heavy isotope enrichments in several rock-forming elements relative to chondrites (assumed to be representative of the planetary precursors) found in various differentiated bodies in the solar system. For example, Paniello et al. (2012) reported excesses in heavy Zn isotopes in lunar rocks compared with terrestrial, martian, and chondritic rocks that they attribute to evaporative loss of Zn during large-sale melting attending the formation of the Moon. Wang and Jacobsen (2016) found an elevated $^{41}K/^{39}K$ for the Moon that they attribute to condensation beyond the Roche limit for the Earth-Moon system just after the giant impact that formed the Moon. This event left the Moon depleted in K relative to Earth. Sossi et al. (2016) used new $^{57}Fe/^{54}Fe$ ratio measurements of SNC meteorites to derive a martian mantle ratio that is lower than the terrestrial values that define BSE. They find a positive correlation between $^{57}Fe/^{54}Fe$ and Fe/Mn (Mn has a lower condensation temperature than Fe) among solar system bodies, including Mars, Vesta (HED meteorite parent body), Earth, Moon, and the angrite parent body. The trend is attributed to evaporation or partial condensation at temperatures of ~ 1300K by these authors. The iron isotope ratios also correlate with bulk silicon isotope ratios, suggesting again that Si isotope ratios in differentiated rocky bodies in the solar system are controlled by volatility. On the other hand, there are experimental data showing that Fe isotopes will also fractionate as the result of high-temperature planetary-scale processes such as partial melting and core formation (Weyer et al. 2005; Dauphas et al. 2009; Williams et al. 2009; Macris et al. 2015; Shahar et al. 2016; Elardo and Shahar 2017). Because of the ambiguities resulting from competing mechanisms for iron isotope fractionation, we focus here on Mg and Si (while mindful that some ambiguity surrounds the fractionation of Si isotopes with sequestration of Si in the core as well).

Pringle et al. (2014) used high $^{29}Si/^{28}Si$ in angrite meteorites relative to chondrites (Figure 1) to suggest that impact-induced evaporation from colliding planetesimals controlled the Si isotopic composition of rocky bodies in the solar system, although Dauphas et al. (2015) argued instead that the angrite data are best explained as the result of isotopic equilibration between SiO gas and forsterite grains in the early solar protoplanetary disk. Both Pringle et al. (2014) and

[1] Hereafter we use the term "planetesimal" for all of these bodies collectively.

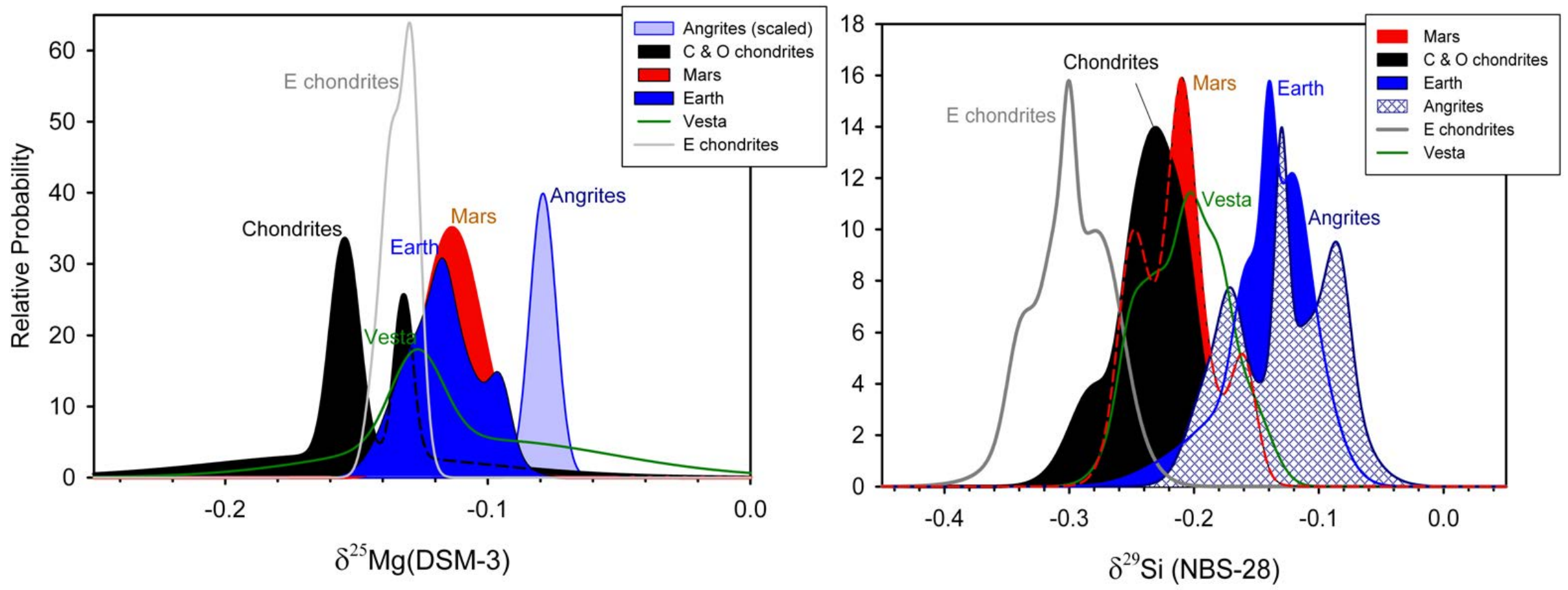

**Figure 1.** Literature compilation of $^{25}Mg/^{24}Mg$ and $^{29}Si/^{28}Si$ for various solar system rocky bodies (Pringle et al. 2013; Savage and Moynier 2013; Zambardi et al. 2013; Hin et al. 2017) represented here as probability density plots. Both isotope ratios are shown as per mil deviations from the standard DSM-3 and NBS-28 materials for Mg and Si, respectively.

Dauphas et al. (2015) used the new angrite Si isotope ratio data to show that sequestration into Earth's core is less likely than volatility to be the primary cause of the high $^{29}Si/^{28}Si$ composition of the bulk silicate Earth (BSE) relative to chondrites.

Young et al. (2009) found that the BSE has higher $^{25}Mg/^{24}Mg$ than chondrites. This early claim has been refined significantly by higher-precision measurements by Hin et al. (2017) (Figure 1, left panel). This latest work suggests that Earth, Mars, Vesta (HED meteorites), and the angrite parent body are all high in $^{25}Mg/^{24}Mg$ relative to chondrites (Figure 1) and the authors attribute this effect to evaporative losses from magma produced by collisions. Hin et al. propose a series of melt-vapor equilibrium events triggered by collisions and melting to explain heavy Mg and Si isotope enrichment of differentiated bodies in the solar system. Norris and Wood (2017) likewise suggest that melt-vapor equilibrium explains the similarity between their experiments on moderately volatile chalcophile elements and the concentrations of these elements in the BSE.

Here we describe the isotopic consequences of evaporation of planetesimal magma oceans. The question we are addressing is whether or not the isotopic compositions of the various solar system bodies for which we have data (Earth, Moon, Mars, Vesta, the angrite parent body) can be reconciled with the hypothesis of planetesimal evaporation, and whether the evaporation could be a natural consequence of planetesimal formation. Hin et al. (2017) considered similar problems. This analysis differs from that work by placing greater emphasis on: 1) the likelihood for isotopic and elemental melt/vapor equilibrium rather than assuming that equilibrium obtains; and 2) the prospects for magma oceans resulting from $^{26}Al$ decay following rapid planetesimal accretion rather than as the result of collisions.

## 2. Planetesimal Magma Oceans

Because of the difficulties of losing vapor from planet-sized bodies due to their strong gravity, attention has focused on volatile losses from smaller bodies on their way to forming planets. Possible candidates include planetesimals with radii up to ~1,000 km that may have grown with the assistance of gas in the protostellar disk leading to rapid formation timescales (Levison et al.

2015; Visser and Ormel 2016). Melt formation by collisions is one popular scenario to consider. Dauphas et al. (2015) calculated that collision-induced vaporization is only significant for targets with masses > 0.2 $M_{\oplus}$. However, higher impact velocities in some N-body calculations (Carter et al. 2015) lead to higher estimates of vapor production by collision (Hin et al. 2017). In any case, nearly instantaneous, complete conversion of liquid or solid to vapor will not itself result in volatility-controlled elemental and isotopic fractionation. Formation of liquid (molten rock) exposed to space for a finite time is required. Numerical simulations of ejecta plumes consider the adiabatic and radiative cooling of the vapor in the plumes (e.g., Richardson et al. 2016), but the temperature evolution of liquids in ejecta plumes will be controlled by the thermal diffusivity and the size scale of the melts as well as by the temperatures and pressures of their environs.

Collisions are not the only mechanism for melting rock in space. Kite et al. (2016) considered evaporation of silicate melt pools on the day-sides of short-period rocky exoplanets and found that the rate of exchange between silicate vapor and the interior of the body depends on the bulk composition of the rocky bodies. In our solar system, early formation of planetesimals heated by the decay of $^{26}$Al may have led to global melting of planetesimals. For example, early accretion of a body with the canonical initial concentration of $^{26}$Al modified by ≤ 1 Myrs of radioactive decay would experience heat production due to radioisotope decay that is ~1,000 times greater per kg than the internal heating induced by tides in present-day Io. This suggests that the melting in these $^{26}$Al-bearing planetesimals far exceeded that which produces the ~50 km thick partial-melt magma ocean and surface magmatism evidenced in and on Io today (Khurana et al. 2011; Tyler et al. 2015; Davies et al. 2018).

In this paper we focus on $^{26}$Al as the heat source for putative magma oceans (here we use the term "magma ocean" in the literal sense of melt exposed on the surface). Parent body mass is a first-order control on the formation of magma oceans, the dynamics of the melt, and the evaporation process. Small bodies lose vapor from evaporating melt while larger bodies can partially retain the evaporation products as a rock vapor atmosphere. Setting the surface escape velocity equal to the mean velocity of a gas molecule results in an equation that serves as a guide for the critical planetesimal radius for retaining gas molecules produced by evaporation:

$$r_{\mathrm{critical}} = \frac{3}{4}\sqrt{\frac{2k_{\mathrm{b}}T}{\pi\, m_{\mathrm{gas}}\,\rho\, G}}\,, \tag{1}$$

where $k_b$ is the Boltzmann constant, $T$ is temperature, $m_{gas}$ is the mass of a gas molecule of interest, $\rho$ is the density of the parent body, and $G$ is the gravitational constant. For SiO gas molecules at ~ 2000 to 1500 K and typical rock densities of 3000 to 4000 kg/m$^3$ the critical radius for gas retention is about 700 km, corresponding to bodies larger than about 0.1% of an Earth mass, 0.001 $M_{\oplus}$, (i.e., ~ ½ the mass of Pluto). We will use this mass as a fiducial model for evaporating planetesimals in what follows, although we consider smaller bodies where nebular gas could have prevented gas loss. Both hydrodynamic escape and Jeans' escape can overcome the escape velocity limit, while repeated collisions could also assist the escape of all elements evaporated from the molten surface (Day and Moynier 2014). The degree of mass bias associated with atmospheric escape depends critically on the mechanism (Zahnle and Kasting 1986; Hunten et al. 1987).

The vigor of convection in the interior of a molten body will control the prospects for a magma ocean. The dynamism of convection can be quantified using the Rayleigh number appropriate for internal heating by radioactive decay (Turcotte and Schubert 2002):

$$Ra = \frac{\rho^2 g \alpha H s^5}{\eta \kappa k}, \tag{2}$$

where $s$ is the radius of the body and $H$ is heat production. Accretion at time zero in the solar system (coinciding with the formation age of calcium-aluminum-rich inclusions) leads to heat production by $^{26}$Al decay of $3\times10^{-7}$ W/kg (for $^{26}Al/^{27}Al = 5 \times10^{-5}$). With typical values for expansivity $\alpha = 3\times10^{-5}$ K$^{-1}$, thermal diffusivity $\kappa = 1\times10^{-6}$ m$^2$ s$^{-1}$, dynamic viscosity $\eta = 1\times10^4$ Pa s, mass density $\rho = 3000$ kg/m$^3$, and the acceleration due to gravity $g = 0.64$ m s$^{-2}$, representing a body with a radius of 700 km and a mass of ½ $M_{\text{Pluto}}$, $Ra = 10^{26}$. The ratio of convective to conductive heat transfer in the body, expressed as the Nusselt number, $Nu$, is ~ $Ra^{1/3}$ (Howard 1966), yielding a value of $10^9$. This analysis suggests that early accretion of planetesimals will have produced bodies with vigorously convecting magma oceans.

The exposure of melt to space on a body heated from within depends on the thickness and stability of the crust comprising the thermal boundary layer between space and melt that in turn scales with $Ra$. By analogy with lava lakes, the surfaces of magma oceans can be envisioned as thin, ductile, visco-elastic crusts constituting the exposed upper boundaries of the convecting melt beneath (Harris 2008; Patrick et al. 2016). Accordingly, these crusts are likely to be dynamic and transient. A rough estimate of the thickness of the crust, $\delta$, can be obtained by considering a thermal steady state such that the radiative flux at the surface balances conductive heat flux through the boundary, yielding (Hin et al. 2017)

$$k\frac{\left(T_{\text{melt}} - T_{\text{eff}}\right)}{\delta} = \sigma T_{\text{eff}}^4, \tag{3}$$

where $\sigma$ is the Stefan-Boltzmann constant, $k$ is the rock thermal conductivity (3 Wm$^{-1}$K$^{-1}$), $T_{\text{melt}}$ is the temperature of the molten rock near the surface, and $T_{\text{eff}}$ is the effective radiative temperature at the surface. Since $T_{\text{eff}}$ is not known *a priori*, a unique solution for the thickness comes from the additional constraint that δ should be smaller with greater vigor of convection. This effect has been quantified for small viscosity contrasts between the melt and warm boundary layer using a scaling in terms of the Rayleigh number for the interior (Solomatov 1995):

$$\delta \sim s\left(Ra\right)^{-1/3}, \tag{4}$$

where again $s$ is the radius of the convecting sphere. By way of example, for our fiducial ½ $M_{\text{Pluto}}$ body, iteration between Equations (3) and (4) with a value for $Ra$ defined by the temperature difference $T_{\text{melt}} - T_{\text{eff}}$ and $T_{\text{melt}} = 1500$ K, leads to $\delta = 0.05$ meters (5 cm) and $T_{\text{eff}} = 890$ K. Utilizing the scaling for a large viscosity contrast between the melt and crust (Solomatov 1995) leads to $\delta \sim 0.03$ meters (3cm) and $T_{\text{eff}} = 1000$ K. These values depend critically on the temperature dependence of $\eta$. With higher values for $Ra$ like that in Equation (2), thinner crusts of ~ 0.1 cm and higher $T_{\text{eff}} > 1400$ K are obtained. Solomatov (2009, and references therein) suggests that the convective velocity $u$ can be estimated from $u \sim 0.6\ (\alpha\ gL\ \sigma T_{\mathrm{eff}}^4/(\rho\ c_{\mathrm{p}}))^{1/3}$ in SI units. For $\alpha = 3\times10^{-5}$ K$^{-1}$, $\rho = 3000$ kg/m$^3$, $g = 0.64$ m s$^{-2}$, scale height $L = 4\times10^5$ meters, $T_{\mathrm{eff}} =$ 1500 K, and specific heat capacity $c_{\mathrm{p}} = 1000$ J kg$^{-1}$ K$^{-1}$, the convective velocity is 0.5 m s$^{-1}$.

We conclude that a dynamic crust << 1 meter thick resulting from energetic convection associated with the magma ocean will be subject to repeated and frequent foundering (Stovall et al. 2009), exposing melt to space more or less continuously.

The timescale for exposure of melt to space is likely to have been short. Melt rising to the surface of a planetesimal accelerates cooling because thermal radiation is proportional to $T_{\mathrm{eff}}^4$ (Equation 3). In order to obtain an estimate of the timescales associated with melting near or at the surface we performed thermal modeling of spherical chondritic planetesimals heated by $^{26}$Al (as well as subordinate $^{60}$Fe and other longer-lived nuclides). The models solve the equations for heat transfer in a body following instantaneous accretion. Details of the basic equations and finite difference solution methods, including the formation of the metal core, are described in detail in Zhou et al. (2013). For this work the isothermal outer boundary condition was replaced with a flux-balance boundary condition based on Equation (3) where conductive heat flux through a thin boundary layer ($4\pi s^2 k(T_{\mathrm{melt}}-T_{\mathrm{eff}})/\delta$) is balanced by the radiative flux at the surface ($4\pi s^2\sigma T_{\mathrm{eff}}^4$). Here $\delta$ is fixed to 0.5 cm based on the analysis above. In addition, thermal diffusivity was enhanced by multiplying $\kappa$ by the Nusselt number where melt fractions exceed 90% as a means of simulating the consequences of convection in the melt below the surface. Numerical stability during cooling limits model values for $Nu$ to $\leq 10^4$ in this calculation although tests with $Nu$ as high as $10^7$ during the magma ocean phase showed similar results to those with lower values.

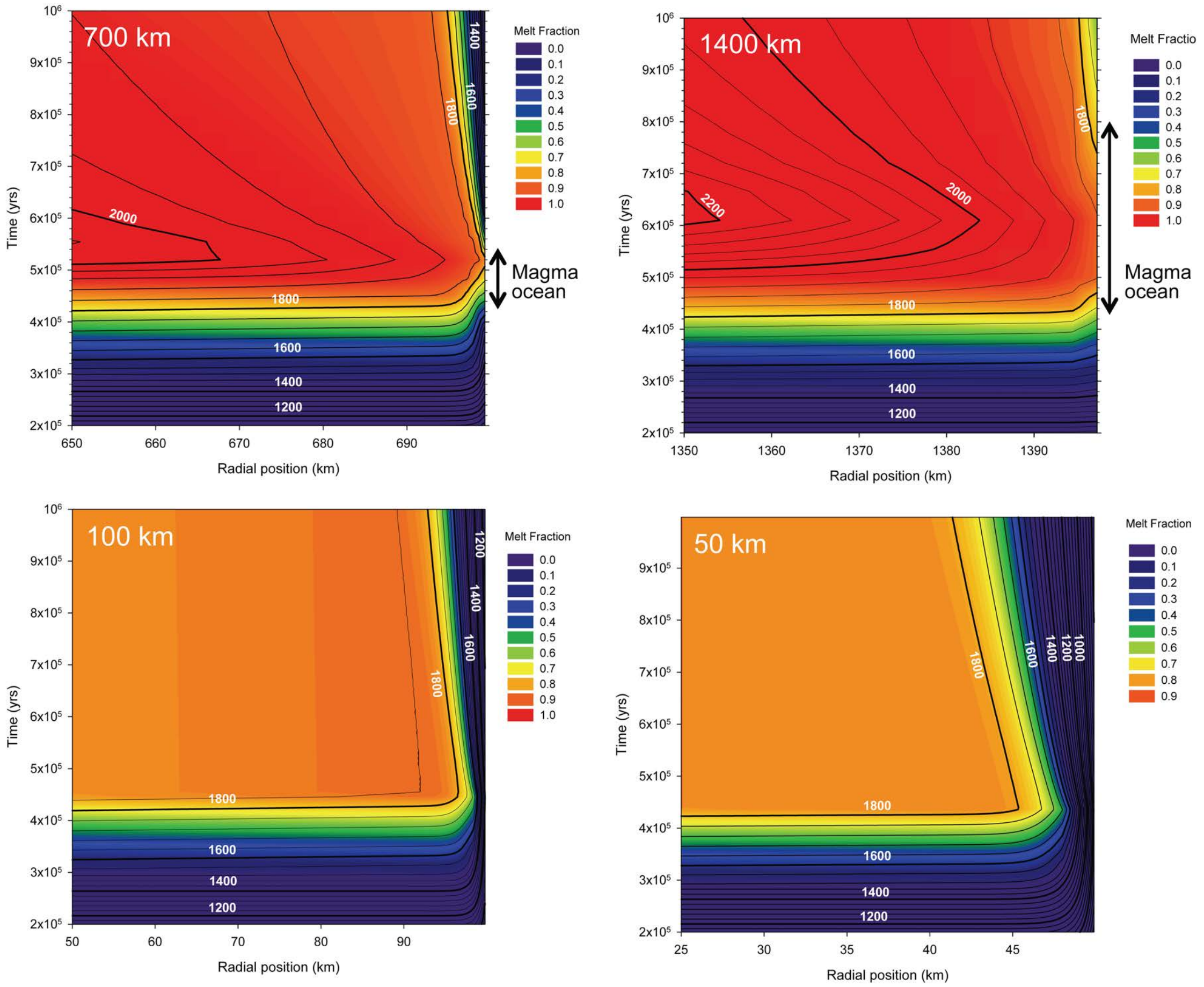

**Figure 2**. Results of thermal models for planetesimals with radii ranging from 50 km to 1400 km that accreted 0.2 Myr post CAI from chondritic material. The abscissa and ordinate in each panel are the radial position in the body and time elapsed after instantaneous accretion, respectively. Colors represent melt fractions while contours show temperature in K. Only the outer-most few kilometers are shown for illustration. A magma ocean exposed to the surface is viable for the two largest bodies depicted here, but not the two smaller bodies. The approximate durations of the surface magma oceans are shown by vertical arrows.

Figure 2 shows the results for bodies ranging in radius from 50 km to 1400 km accreted 0.2 Myr post CAI. The prospects for a magma ocean exposed at the surface depend on accretion time and size of the body. We find that only bodies formed within ~$10^5$ years of time zero in the solar system and ~600 km in radius or larger experience a magma ocean exposed to the surface. The models serve to estimate the approximate timescale for a magma ocean phase if it were to occur, as our calculations suggest. For our fiducial ½ $M_{\mathrm{Pluto}}$ body (700 km radius), the potential magma ocean phase persists for ~ $10^5$ years (Figure 2). The timescale for surface melting for a body twice this size, representing the upper mass limit for planetesimals considered here, is ~ $3\times10^5$ years or less (Figure 2). Melting just below or at the surface ends by a combination of rapid cooling and the decrease in $^{26}$Al decay with time; the abundance of $^{26}$Al decreases by half after $7\times10^5$ years. Our $10^5$ year duration for a magma ocean is broadly consistent with the longevity of a Vestan near-surface magma ocean suggested by the models presented by Neumann et al. (2014) .

The timescale for accreting plantesimals of the size required for magma oceans is subject for debate. We note here that the pebble accretion models of Visser and Ormel (2016) suggest that planetesimals with radii > ~ 300 km may grow in $10^5$ years, consistent with the assumed accretion time of $2\times10^5$ yrs in Figure 2. Smaller bodies take longer to form in their models. Bodies as large as 1000 km could grow in as little as $10^4$ years.

## 3. Evaporation

### *3.1 Evaporative fluxes*

Evaporation depends on the difference between the saturation (i.e., equilibrium) vapor pressure and the extant vapor pressure surrounding the evaporating melt. More specifically, the evaporative flux of a species $i$ from a melt is given by the Hertz-Knudsen equation:

$$J_{i,\,\mathrm{net}} = \frac{\gamma_i\left(P_{i,\,\mathrm{sat}} - P_i\right)}{\sqrt{2\pi m_i RT}} \quad , \tag{5}$$

where $P_{i,\mathrm{sat}}$ is the equilibrium vapor pressure of gas species $i$, $P_i$ is the actual vapor pressure, $m_i$ is the molecular mass of species $i$, $\gamma_i$ is the dimensionless evaporation coefficient for $i$, $R$ is the ideal gas constant, $T$ is temperature, and $J_{i,\mathrm{net}}$ is the net evaporative flux of gas species $i$. The $\gamma_i$ parameter that appears in the evaporation equations is akin to the sticking coefficient for adsorption. It cannot in general be evaluated from first principles. Rather, it must be measured. In this way it plays a role analogous to the activity coefficient in thermodynamic chemical potentials.

A crucial aspect controlling evaporation behavior is the size of the evaporating body relative to the mean free path of the surrounding gas because the size of the body affects $P_i$. The critical radius of an evaporating "droplet" (in this case the droplet is a planetesimal) for which the vapor pressure for an evaporating species surrounding the droplet is half the saturation pressure is $r_{\text{critical, sat}} \sim (4/3)\ l/\gamma$ where $l$ is the mean free path of the gas and $\gamma_i$ is the empirical evaporation coefficient that appears in Equation (5) (Kozyrev and Sitnikov 2001). This illustrates that larger bodies approach saturation at lower pressures (larger $l$) than smaller bodies because the return flux of gas due to collisions with surrounding gas molecules is greater for larger surface areas (Figure 3). For example, millimeter-sized bodies like molten chondrules will reach saturation only where the enveloping gas pressure is on the order of 1 bar ($10^5$ Pa) whereas asteroidal-sized bodies will approach saturation at considerably lower pressures (see below).

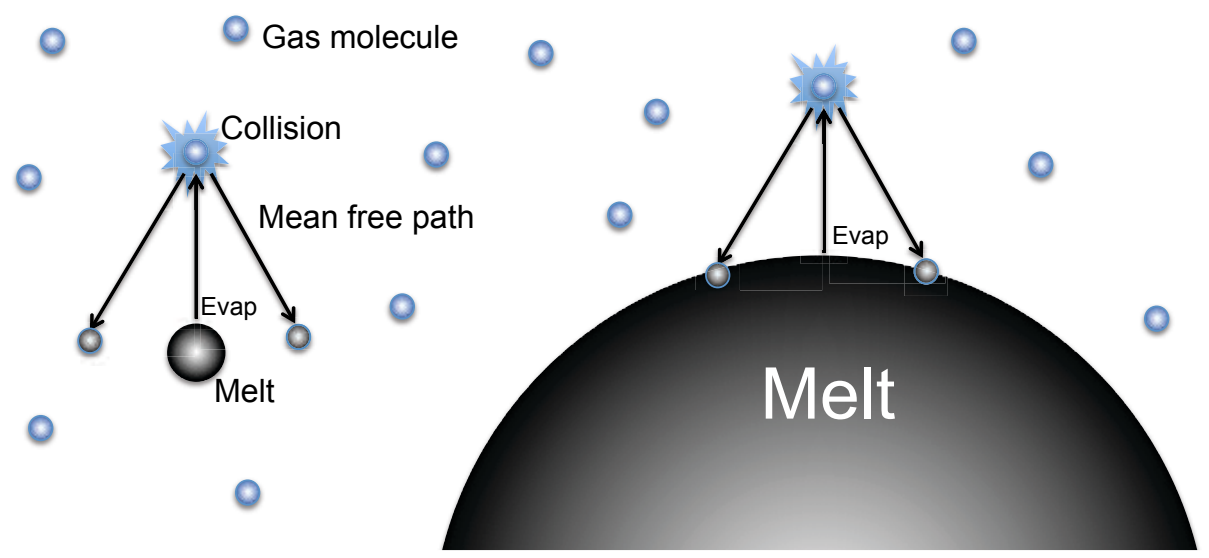

**Figure 3**. Schematic illustration of the effect of droplet size on return flux of gas. Evaporation is depicted by the upward vertical arrows. Evaporated molecules (small black spheres) collide with surrounding gas molecules (blue spheres). Droplets that are small in diameter compared with the mean free path (left side) capture less of the return flux than droplets that are large relative to the mean free path (right side).

Equation (5) shows that the net flux of a species $i$ from the surface of an evaporating body, $J_{i,\text{net}}$, is the difference between the the evaporative flux, $J_{i,\text{evap}}$ that is proportional to the equilibrium vapor pressure $P_{i,\text{sat}}$ and the return flux $J_{i,\text{return}}$ that is proportional to the total local partial pressure $P_i$. The diffusion-limited return flux leads to a pressure build up adjacent the surface of the evaporating body that can be described using (Carslaw and Jaeger 1959):

$$P_i(r_+ = s,t) = RT\frac{s}{D_i}J_{i,\text{net}}\left[1 - e^{\xi}\,\text{erfc}\left(\sqrt{\xi}\right)\right] \quad , \tag{6}$$

where $P_i$ ($r_+ = s$, $t$) is the pressure of gas $i$ immediately above the liquid surface, $r_+$ is the radial distance above the surface of the spherical evaporating body of radius $s$, $D_i$ is the gas-phase diffusion coefficient for the species $i$, and $\xi = tD_i/s^2$ is the dimensionless elapsed time from the start of evaporation. Equation (6) provides a means of quantifying saturation as $P_i/P_{i,\text{sat}}$.. It is straightforward to show that $J_{i,\text{net}}/J_{i,\text{evap}} = 1/(1+J_{i,\text{return}}/J_{i,\text{net}})$. Solving this relationship for the net flux of species $i$ with substitution of Equation (6) into Equation (5) yields (e.g., Richter et al. 2002)

$$J_{i,\text{net}} = \frac{J_{i,\text{evap}}}{1 + \dfrac{\gamma_i RT}{\sqrt{2\pi m_i RT}}\dfrac{s}{D_i}\left[1 - e^{\xi}\,\text{erfc}\left(\sqrt{\xi}\right)\right]} \tag{7}$$

from which it is clear that for a given mass and evaporation coefficient, the net evaporative flux depends on the radius of the evaporating body, $s$, and the far-field ambient pressure $P_\infty$ since $D_i = \left(1/(3\sigma P_\infty\sqrt{\mu}\,)\right)(k_bT)^{3/2}$, where $\sigma$ is the collisional cross section and $\mu$ is the reduced mass for the species of interest and its collision partners. Where the far-field pressure is effectively zero, $D_i \rightarrow \infty$ and $J_{i,\,\mathrm{net}} = J_{i,\,\mathrm{evap}}$. Figure 4 shows the ratio of net flux to free evaporation flux, $J_{i,\mathrm{net}}/J_{i,\mathrm{evap}}$, for $^{24}$Mg gas evaporating into a background pressure of $10^{-8}$ bar ($10^{-3}$ Pa) as a function of the radius of the evaporating body. The plot illustrates that at this pressure bodies on the order of 10 km in radius experience free evaporation where $J_{i,\mathrm{net}}/J_{i,\mathrm{evap}} \sim 1$ while bodies of order 1000 km in radius approach equilibrium where $J_{i,\mathrm{net}}/J_{i,\mathrm{evap}} << 1$.

### *3.2 Isotopic effects of evaporation*

The isotopic effects of evaporation and condensation must be included in order to arrive at the net effect of both processes operating simultaneously. The condensation isotope fractionation factor, $\alpha_{\mathrm{cond}}$, is related to the equilibrium and evaporation fractionation factors by the law of mass action:

$$\alpha_{\mathrm{eq}} = \frac{\alpha_{\mathrm{cond}}}{\alpha_{\mathrm{evap}}} \quad . \tag{8}$$

Here $\alpha_{\mathrm{eq}}$ is the atomic ratio of heavy-to-light isotopes in the melt divided by that of the vapor at equilibrium, or $((n'_{i,\mathrm{melt}}/n_{i,\mathrm{melt}})/(n'_{i,\mathrm{vapor}}/n_{i,\mathrm{vapor}}))_{\mathrm{eq}}$ where $n_i$ is the number density for the species $i$ and where the superscript prime signifies the heavy isotope. The evaporation fractionation factor $\alpha_{\mathrm{evap}}$ is the isotope ratio of the vapor divided by that of the melt, or $((n'_{i,\mathrm{vapor}}/n_{i,\mathrm{vapor}})/(n'_{i,\mathrm{melt}}/n_{i,\mathrm{melt}}))_{\mathrm{evap}}$, only where now the isotope or isotopologue ratio in the numerator is determined by the kinetics of evaporation. Equation (8) is important because the return flux of the gaseous heavy isotope or isotopologue to the melt from the vapor is related to that of the light isotopic species by $\alpha_{\mathrm{cond}}$. While we can derive the melt/vapor equilibrium fractionation factor, $\alpha_{\mathrm{eq}}$, from first principles and $\alpha_{\mathrm{evap}}$ can be obtained relatively easily from laboratory experiments, $\alpha_{\mathrm{cond}}$ is less tractable both from theory and from experiments. Using Equation (8), however, we can calculate the condensation fractionation factor as $\alpha_{\mathrm{cond}} = \alpha_{\mathrm{eq}}\, \alpha_{\mathrm{evap}}$. The bond energy effect of condensation applies to the return flux from the gas phase to the melt such that $J^{*\prime}_{i,\mathrm{return}} = \alpha_{\mathrm{eq}}\alpha_{\mathrm{evap}}\, J'_{i,\mathrm{return}}$. Here we have used Equation (8) for the condensation fractionation factor, the prime symbol to denote the heavy isotopic species as usual, and the star symbol to signify the return flux corrected for the bond energy effects of condensation. We can then write the ratio of the net and evaporative fluxes as before but for the heavy isotopic species, yielding $J^{*\prime}_{i,\mathrm{net}}/J'_{i,\mathrm{evap}} = 1/(1+ \alpha_{\mathrm{eq}}\alpha_{\mathrm{evap}}\, J'_{i,\mathrm{return}}/J'_{i,\mathrm{net}})$. By taking the ratio of the net evaporative fluxes (Equation 7) for the heavy and light isotopic species relative to the abundances of the isotopes of interest in the melt phase, inclusive of bond energy effects of condensation, we arrive at the net isotope fractionation factor $\alpha_{\text{net}}$:

$$\alpha_{\text{net}} = \frac{\left(\dfrac{J^{*\prime}_{i,\text{net}}}{J_{i,\text{net}}}\right)}{\left(\dfrac{n'_{i,\text{melt}}}{n_{i,\text{melt}}}\right)} = \alpha_{\text{evap}} \frac{1 + \dfrac{\gamma_i RT}{\sqrt{2\pi m_i RT}} \dfrac{s}{D_i}\left[1 - e^{\xi}\,\text{erfc}\left(\sqrt{\xi}\right)\right]}{1 + \dfrac{\alpha_{\text{eq}}\alpha_{\text{evap}}\gamma_i RT}{\sqrt{2\pi m_i' RT}} \dfrac{s}{D_i'}\left[1 - e^{\xi'}\,\text{erfc}\left(\sqrt{\xi'}\right)\right]} \quad . \tag{9}$$

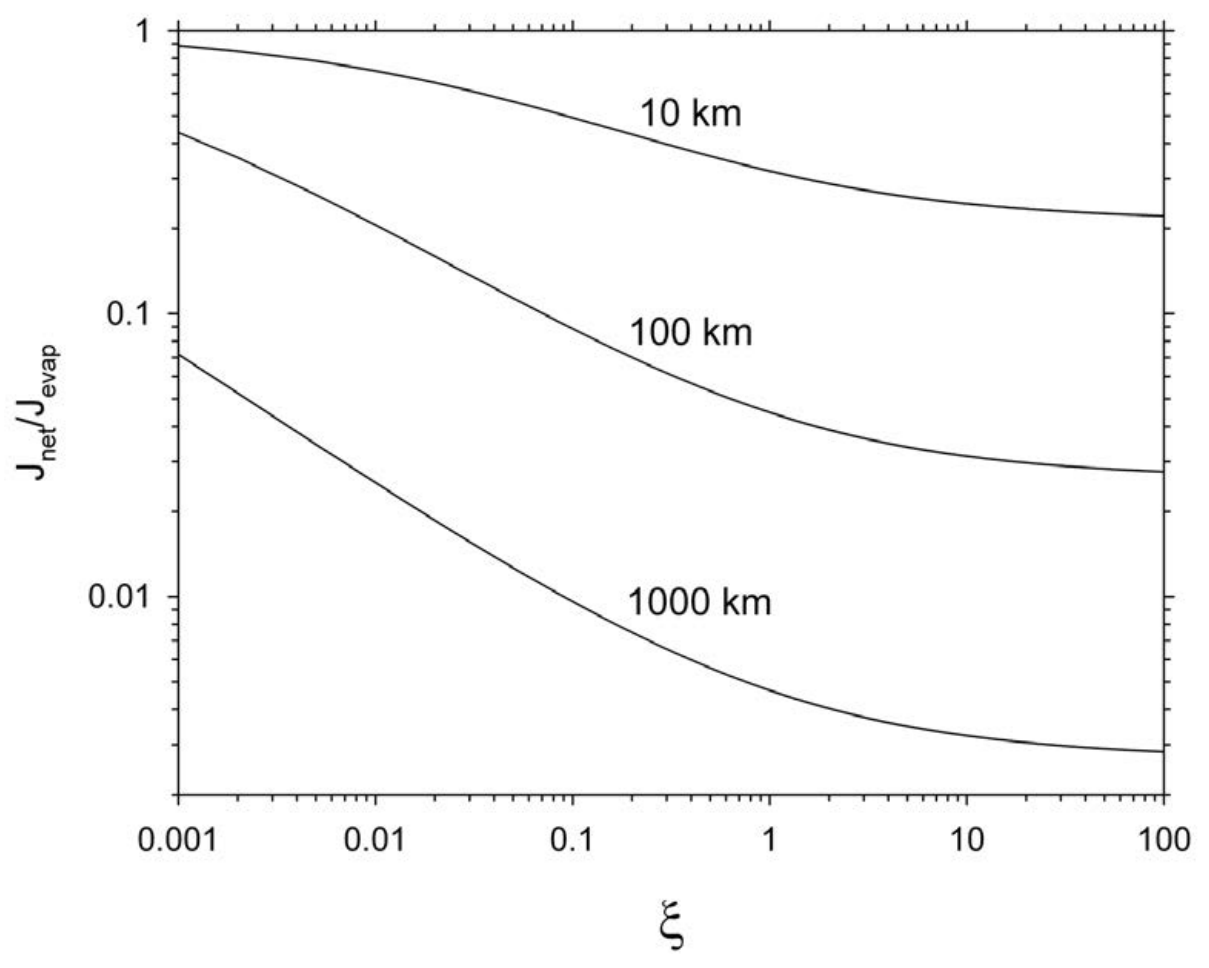

**Figure 4.** Plot of the ratio of net flux to evaporation flux vs dimensionless time $\xi$ for an evaporating spherical body according to Equation (7) with $D_{\mathrm{Mg}}$ evaluated at 2000 K and $10^{-8}$ bar total pressure. Contours for three planetesimal radii, $s$, are shown.

The reader can verify that in the limit of long relative timescales (large $\xi$, where $\xi \sim t\ 10^{-8}$ for $t$ in seconds in our application) and a high return flux, $J^{*\prime}_{i,\text{net}}/J_{i,\text{net}}$ as given by Equation (9) approaches $(n'_{i,\text{melt}}/n_{i,\text{melt}})\alpha_{\text{evap}}\sqrt{\mu'/\mu}\sqrt{\mu/\mu'}/\alpha_{\text{eq}}\alpha_{\text{evap}} = (n'_{i,\text{melt}}/n_{i,\text{melt}})/\alpha_{\text{eq}}$, which is the composition of the vapor in equilibrium with the melt; the limit of high return flux is equilibrium isotope partitioning between gas and melt. We note that in arriving at this asymptotic evaluation of (9) we have invoked the requirement that the masses that appear in Equation (9) should be the same reduced masses, $\mu$ and $\mu'$, as those that apply to the diffusivities in the gas phase since both refer to transport through the gas. These will usually be the reduced masses defined by the diffusing species *i* and the background gas.

In order to evaluate Equation (9), values for $\alpha_{\text{eq}}$ and $\alpha_{\text{evap}}$ are required. In this study we use the experimentally-determined free evaporation fractionation factors for calcium-aluminum-rich inclusion melts (instances of CaO-MgO-$Al_2O_3$-$SiO_2$, or CMAS, melts). The $\alpha_{\text{evap}}$ values for $^{25}$Mg/$^{24}$Mg and $^{29}$SiO/$^{28}$SiO of 0.9869 and 0.9898 reproduce available experimental data (Shahar and Young 2007). Equilibrium fractionation factors have not been measured between vapor and silicate melt, but reasonable values can be obtained from ratios of reduced partition function ratios for forsterite (representing melt) and the relevant gas species. Magnesium evaporates as the atomic species Mg so for $^{25}$Mg/$^{24}$Mg the reduced partition function ratio for forsterite is the equilibrium melt/vapor fractionation factor. Silicon evaporates as SiO so the equilibrium fractionation factor is the ratio of the reduced partition function ratios for forsterite and SiO. For the Mg equilibrium isotope fractionation between melt and vapor we have $\alpha_{\text{eq}}$ ($^{25}$Mg/$^{24}$Mg) melt/vapor = exp[(0.520/1000) $2.107\times10^6/T^2$] where $T$ is in K (Schauble 2011). For Si we obtain in this study $\alpha_{\text{eq}}$ ($^{29}$Si/$^{28}$Si) melt/vapor = exp[(1/1000) $2.466\times10^6/T^2$] using methods similar to those described by Shahar et. al. (2009) and Ziegler et al. (2010). At 2000 K, a rough approximation for the upper end of the melt temperature range, the equilibrium melt/vapor fractionation factors for $^{25}$Mg/$^{24}$Mg and $^{29}$Si/$^{28}$Si are 1.000274 and 1.000617, respectively.

## 4. Rock Vapor Atmospheres

### *4.1 Evaporative source*

Smaller planetesimals will lose gas readily upon evaporation because thermal velocities exceed escape velocities unless the bodies are enveloped by the $H_2$-rich gas of the protoplanetary disk (see below). For planetesimals massive enough to retain hot gas (> ~ $5 \times 10^{21}$ kg, ~ $1/2M_{\text{Pluto}}$), one can show that a steady-state rock vapor atmosphere forms within minutes to hours and results from a balance between rates of magma evaporation (Equation 7) and atmospheric escape. In order to quantify the pressure of the rock vapor atmosphere surrounding an evaporating planetesimal, evaporative fluxes from the melts are required.

We understand a great deal about the evaporative fluxes emanating from CMAS melts both from theory and experiments (e.g., Davis et al. 1990; Young et al. 1998; Grossman et al. 2000; Alexander 2001; Richter et al. 2002; Richter 2004; Richter et al. 2007; Shahar and Young 2007; Grossman et al. 2008; Knight et al. 2009). There are also studies of the physical chemistry of chondritic melt evaporation (Floss et al. 1996; Alexander 2001; Fedkin et al. 2006), although they are fewer in number. The equilibrium vapor pressures that drive free evaporation can be parameterized within a set of reference evaporative fluxes for CMAS melts described previously (Grossman et al. 2000; Richter et al. 2002; Shahar and Young 2007). The reference fluxes permit calculation of evaporative fluxes with adjustments for melt compositions and ambient gas compositions. Here we have modified the calculation method described by Shahar and Young to include non-ideal mixing for $SiO_2$ in the melt (ideal mixing is used for the other components MgO, CaO and $Al_2O_3$) in order to expand the applicable range in composition space. In these calculations, the evaporative flux of SiO from a melt with a specified mole fraction of $SiO_2$, $x_{SiO2}$, is related to the flux from the reference CMAS material by the expression

$$J_{SiO} = J_{SiO}^{ref} \frac{x_{SiO_2}}{x_{SiO_2}^{ref}} \exp\left[\frac{-2Wx_{SiO_2}}{RT}\left(\frac{x_{SiO_2}}{x_{SiO_2}^{ref}} - 1\right)\right] , \tag{10}$$

where $W$ is the symmetrical interaction parameter, the superscript "ref" refers to the reference fluxes reported in the literature, and higher order products of mole fractions have been ignored in the activity coefficients. In the context of a regular solution model, we have determined empirically from the CMAS experimental data that to a good approximation, $-2Wx_{SiO_2}/(RT) = 8$. The IDL+Fortran code that includes the activity model in Equation (10) faithfully reproduces the bulk chemistry and isotopic compositions of CMAS evaporative residues (Figure 5, see §5 for a complete description of the calculations). When applied to chondrite-like bulk compositions renormalized to exclude Fe and alkali elements, the model works within uncertainties, although the existing experimental data are not sufficient to test the results in detail (Figure 5).

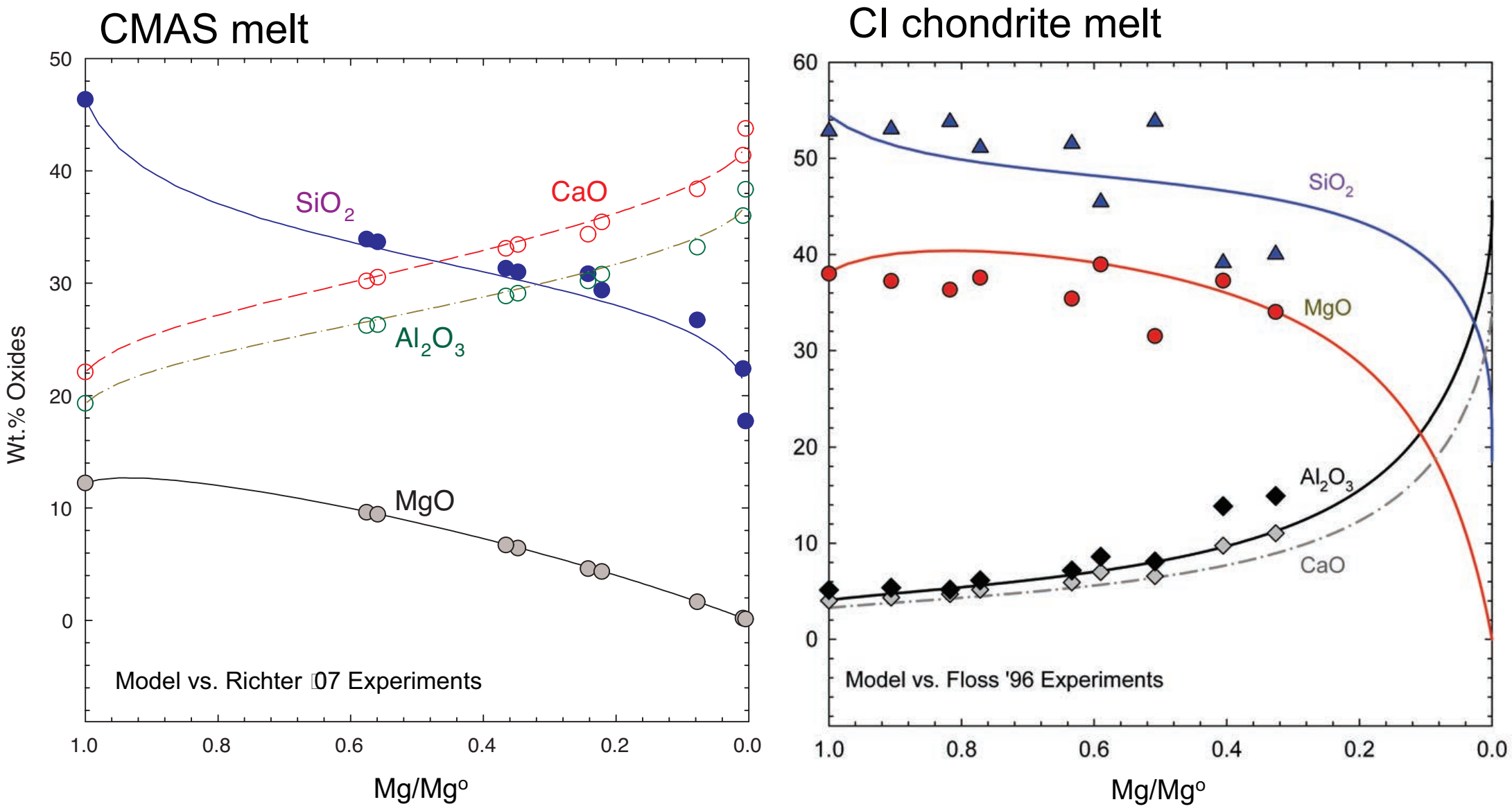

**Figure 5**. Left: UCLA model for evaporation of a CMAS melt compared with experiments at 1675K for the same bulk composition (Richter et al. 2007). Right: UCLA model applied to evaporation of a carbonaceous chondrite bulk composition compared with experiments on evaporation of the Allende chondrite (Floss et al. 1996) projected into the iron-free CMAS composition space. The abscissa is the mass of Mg relative to the initial mass of Mg, Mg°.

*4.2 Atmospheric escape*

In order to estimate atmospheric escape rates we need to specify the nature of the rock-vapor atmosphere. Like other atmospheres, the atmosphere created by evaporation can have three distinct layers, a well-mixed convective layer (troposphere), a radiative layer (stratosphere), and a collisionless exosphere, with relative depths controlled by the mass and surface temperature of the parent body (in the case of a magma ocean surface, energy from the star is negligible). Conduction at the hot magma surface will heat the base of the atmosphere, inducing a sharp temperature gradient that may lead to convection. In the convecting layer temperature and pressure will vary in a manner asymptotic to adiabatic, such that

$$\frac{d\ln(T)}{d\ln(P)} = \frac{R}{C_p} \tag{11}$$

where the ratio of the ideal gas constant to the isobaric heat capacity on the right-hand side of (11) is 2/7 for an ideal diatomic gas (e.g., ideal behavior of SiO). Assuming a motionless or slowly-moving gas, hydrostatic equilibrium applies and we can further specify that

$$\frac{dT}{dz} = -\frac{\gamma - 1}{\gamma}\frac{m_{\text{gas}}\, g}{R}\ , \tag{12}$$

where $g$ is the acceleration due to gravity, $\gamma$ is the gas adiabatic constant, or isobaric to isochoric heat capacity ratio $Cp/Cv$ (7/5 for ideal diatomic), and $z$ is altitude relative to the melt surface. This layer is characterized by decreases in pressure and temperature with altitude. Above the troposphere heat transfer is controlled by radiation rather than convection. In this "stratosphere" where radiative and hydrostatic equilibrium obtain, temperature varies according to (de Pater and Lissauer 2006)

$$\frac{dT}{dz} = -\frac{3}{16}\frac{\alpha_R\, \rho_{gas}}{T^3}T_{\text{T}}^4\ , \tag{13}$$

where $\rho_{\text{gas}}$ is the mass density for the gas, $\alpha_R$ is the mean Rosseland opacity of the gas (<< 1 m$^2$ kg$^{-1}$ in this application) and $T_{\text{T}}$ is the effective radiation temperature at the tropopause. The optical depths in the present case are << 1 and usually < 0.1, meaning that the radiation temperature is close to that of the surface.

For an optically-thin atmosphere, the position of the tropopause is defined by equating the adiabatic temperature at the top of the troposphere with the temperature of the stratosphere, yielding (Pierrehumbert 2010)

$$\frac{P_{\text{Tropopause}}}{P_{\text{S}}} = 2^{-C_p/(4R)} \tag{14}$$

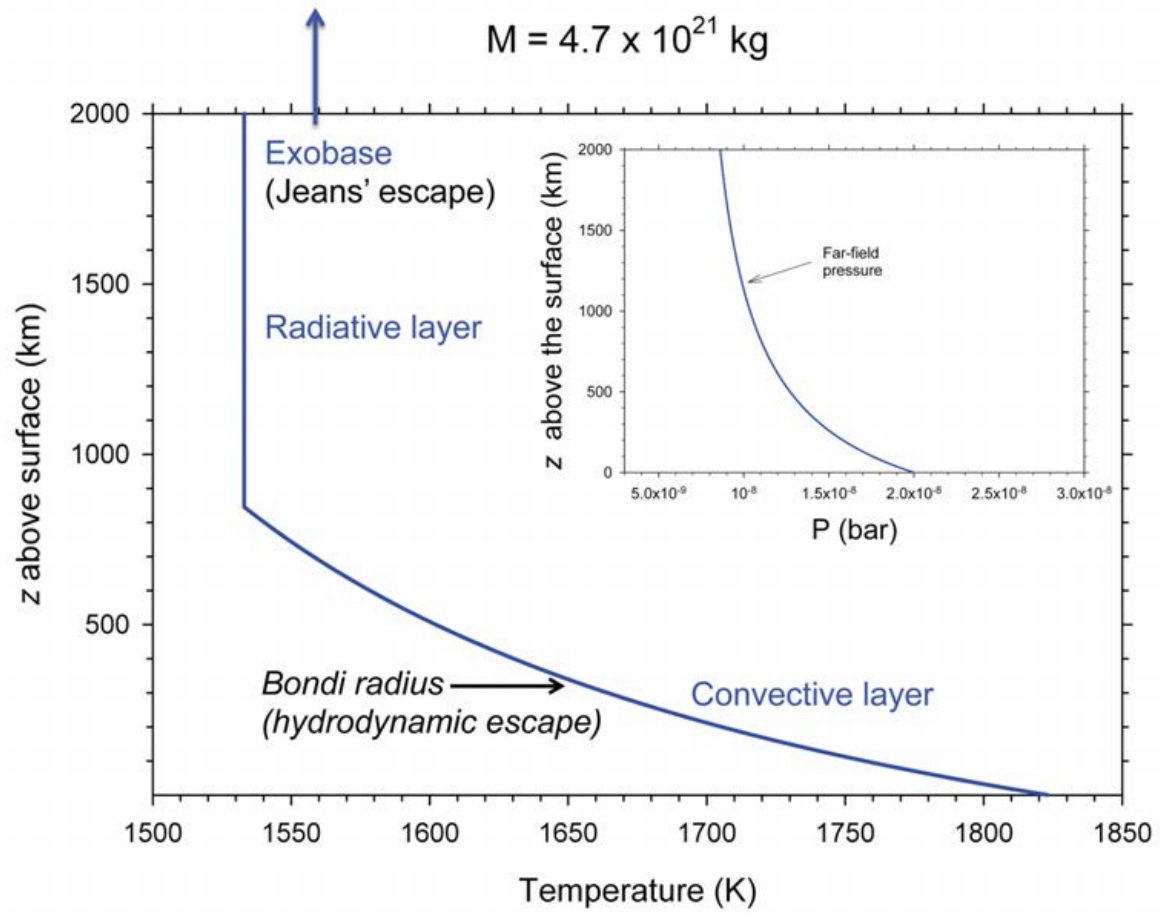

**Figure 6**. Rock vapor atmospheric temperature and pressure as a function of altitude above the melt surface $z$ based on the calculations described in the text for a body that has a mass ~ ½ that of Pluto. Above the Bondi radius that is within the troposphere in this example, the atmosphere would no longer be bound, but the complete profile with $z$ is shown for reference.

where the exponent on the right is –7/8 for an ideal diatomic gas and so $P_{\text{Tropopause}} \sim 0.6\ P_S$ where $P_S$ is the pressure at the surface. Pressure is relatively constant above the tropopause. The altitude of the exobase can be estimated using $l(r)/H(r) = 1$, where $H(r)$, the scale height at radius $r$, is related to the scale height at the surface by $H(r) = H_S\ (r/s)^2$ and the mean free path $l$ is ~ $1/(n_{ex}\ \sigma_{coll})$ where $n_{ex}$ and $\sigma_{coll}$ are the number density at the exobase and collisional cross sections, respectively. We use $\sigma_{coll} = 3 \times 10^{-20}$ m$^2$. Equations (11) through (14) provide the basis for the simplified model rock vapor atmosphere for our fiducial ½ $M_{\text{Pluto}}$ magma ocean body shown in Figure 6. The temperature drops from the near-surface values of ca. 1850 K (melt temperature, e.g. Figure 2) to ~ 1530 K at the tropopause ~800 km above the surface. The temperature remains nearly isothermal for hundreds of kilometers above the tropopause.

Two escape mechanisms are feasible for bodies large enough to retain gas, hydrodynamic escape and Jeans' escape. Hydrodynamic escape will occur at the Bondi radius, $r_B = 2GM/C_S^2$, where the sound speed, $C_S = \sqrt{\gamma k_b T / m_{gas}}$ , equals the escape velocity. For the ½ $M_{\text{Pluto}}$ case in Figure 6, $r_B$ = 1013 km, equivalent to $z$ = 313 km above the melt surface. Purely thermal escape, or Jeans' escape, occurs in the region of the exobase.

The relative efficacies of these two escape mechanisms are evaluated using the escape parameter, $\lambda = (GMm_{gas}/r)/(k_bT(r))$, where $M$ is the mass of the body and $T(r)$ is the temperature at radius $r$ from the center of the body. The escape parameter represents the ratio of gravitational energy to thermal energy; larger values for $\lambda$ favor Jeans' escape over hydrodynamic escape because in this case, gas is trapped in the gravity well of the body until it reaches the rarefied exobase. Volkov et al. (2011) used the escape parameter evaluated at the surface of the parent body, $\lambda_O$, to characterize the relative importance of the two escape process in circumstances where the atmosphere is relatively thin and heated from below, as in this application. Their work verifies that hydrodynamic escape dominates for $\lambda_O < \sim 2$ while Jeans' escape dominates for $\lambda_O$> ~3 to 4. Figure 7 shows contours for $\lambda_O$ in terms of parent body mass and surface temperature. Figure 7 suggests that on planetary embryos where $M \geq \sim M_{Mars}$, evaporation from magma oceans ($T_S$ > 1500 K) will be balanced by Jeans' escape while for planetesimals where $M \leq 10^{22}$ kg (~ mass of Neptune's moon Triton or less), evaporation will be balanced against hydrodynamic escape.

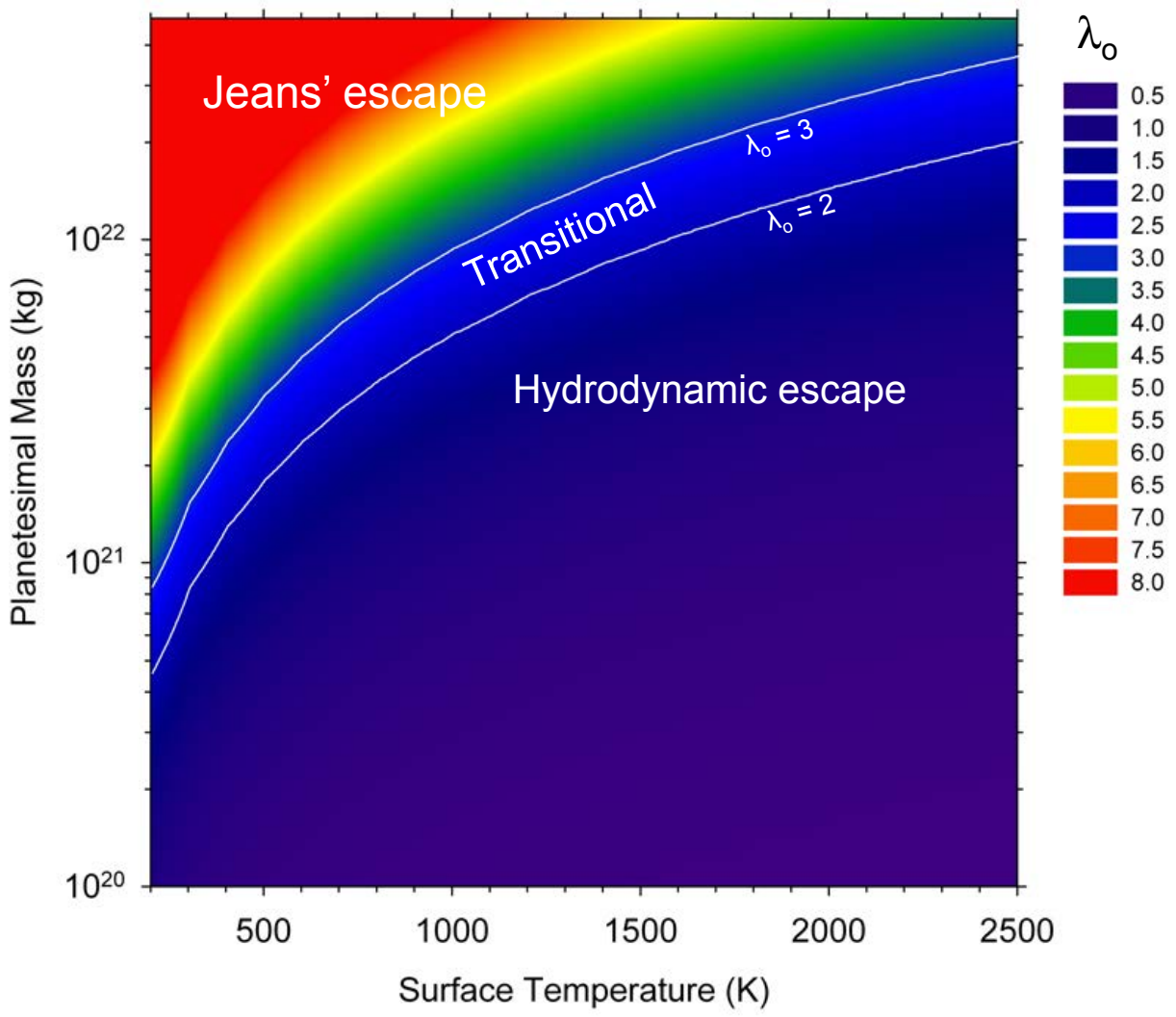

**Figure 7.** Contours of the escape parameter evaluated at the melt surface, $\lambda_O$, of bodies with different masses and surface temperatures.

The surface-integrated hydrodynamic molecular escape mass flux (molecules/s) is

$$\theta_{\text{hydro}} = 4\pi r_B^2 C_S(r_B)\left(\frac{P(r_B)}{k_b T}\right) \tag{15}$$

where $C_S(r)$ is evaluated at $r = r_B$ and $P(r_B)/(k_b T)$ is the number density at that altitude. The surface-integrated molecular flux due to Jeans' escape is (Hunten 1973; Pierrehumbert 2010)

$$\theta_{\text{Jeans'}} = 4\pi r_{\text{Ex}}^2 \omega_J n_{\text{Ex}} \tag{16}$$

where $n_{Ex}$ is the molecule number density at the exobase, $r_{Ex}$ is the radius of the exobase relative to the center of the body, and $\omega_J$ is the Jeans' velocity given by

$$\omega_J = \frac{1}{2\sqrt{\pi}}(1+\lambda(r_{\text{Ex}}))\,\exp\left(-\lambda(r_{\text{Ex}})\right)\sqrt{\frac{2k_b T_{\text{Ex}}}{m_{\text{gas}}}}\ , \tag{17}$$

and where $\lambda(r_{Ex})$ is the escape parameter evaluated at the radius of the exobase. Both surface-integrated fluxes due to atmospheric escape increase with greater pressures at the mass-loss radii. This is in contrast to the rates of evaporation at the melt surface that decrease with far-field pressure because greater far-field pressures decrease gas diffusivities and smaller diffusivities increase partial pressures of evaporating species at the surface of the melt (Equation 6). The opposing effects of pressure on evaporation and escape will result in a steady-state rock vapor atmosphere even where no enveloping nebular gas is present.

We quantified this effect for our fiducial ½ $M_{Pluto}$ body. For these calculations we used the melt composition of an EL chondrite excluding iron (i.e., E chondrite projected into CMAS composition space). The cross-over far-field pressure (in this context the far-field pressure means the pressure at the Bondi radius), where evaporative and hydrodynamic escape are equal, defines the steady-state pressure. In our example, this pressure is

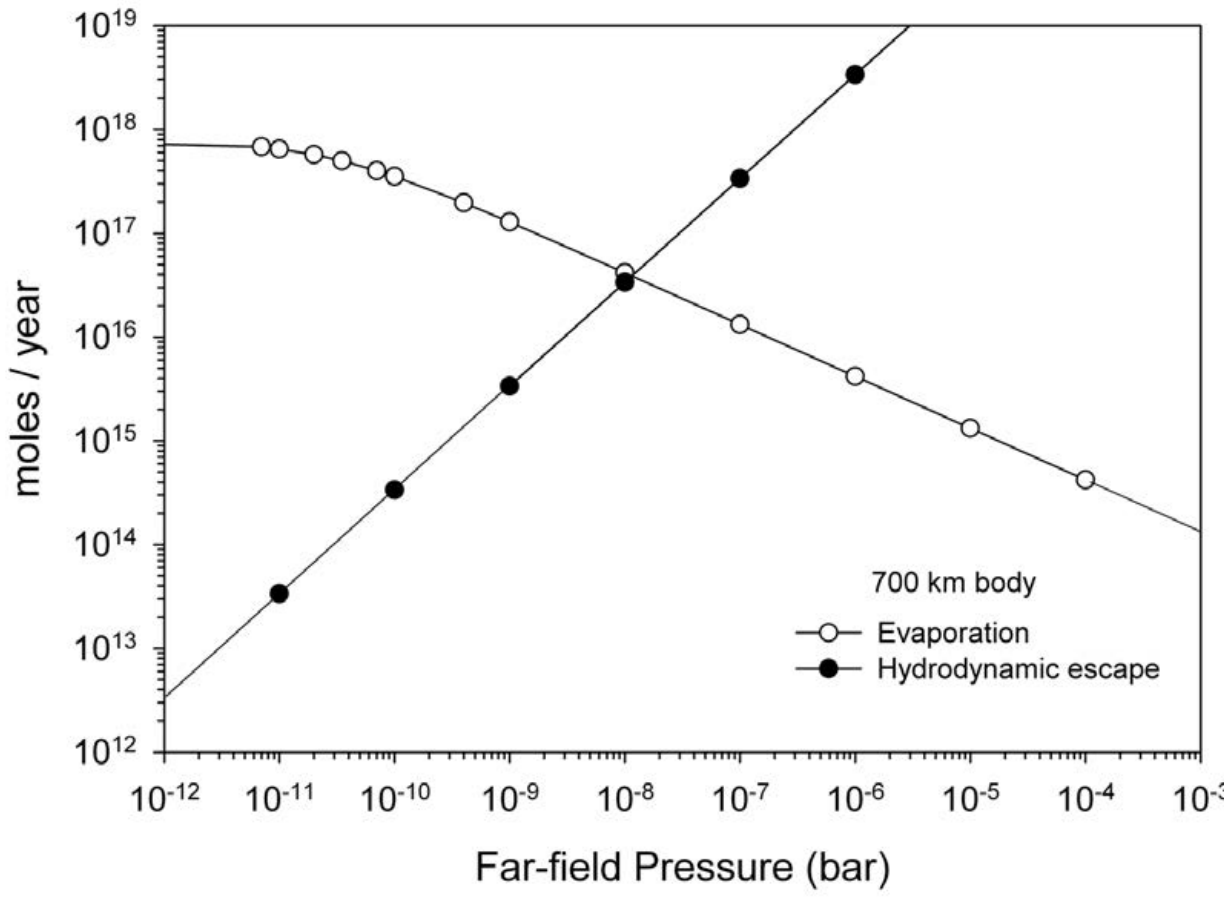

**Figure 8**. Comparison of the surface-integrated fluxes due to evaporation of a molten body half the mass of Pluto composed of E chondrite, excluding iron (E chondrite bulk composition projected into the CMAS composition space) and hydrodynamic escape of the resulting atmosphere. The cross-over pressure represents the steady-state pressure of the rock-vapor atmosphere surrounding the body.

~$10^{-8}$ bar ($10^{-3}$ Pa) (Figure 8). For comparison, a pressure of 1 bar ($10^5$ Pa) or greater would be required for saturation of a cm-sized droplet of evaporating melt.

Using Equation (6) one finds that the vapor pressure buildup immediately adjacent the surface of the evaporating melt results in a pressure of $10^{-4}$ bar at that location (several orders of magnitude greater than the far-field pressure), corresponding to ~ 95% saturation. The isotopic consequences of this build up of gas at the evaporating surface are described by Equation (9). The net evaporation fractionation factors for Mg and Si isotopes are shown as a function of far-field pressure in Figure 9. As pressure increases, the net fractionation factors approach equilibrium (near unity). At the steady-state gas pressure, the saturation of Mg and SiO of ~ 95% for the ½ $M_{\mathrm{Pluto}}$ body means that the net isotope fractionation factors for Mg and Si are 95% towards equilibrium (Figure 9).

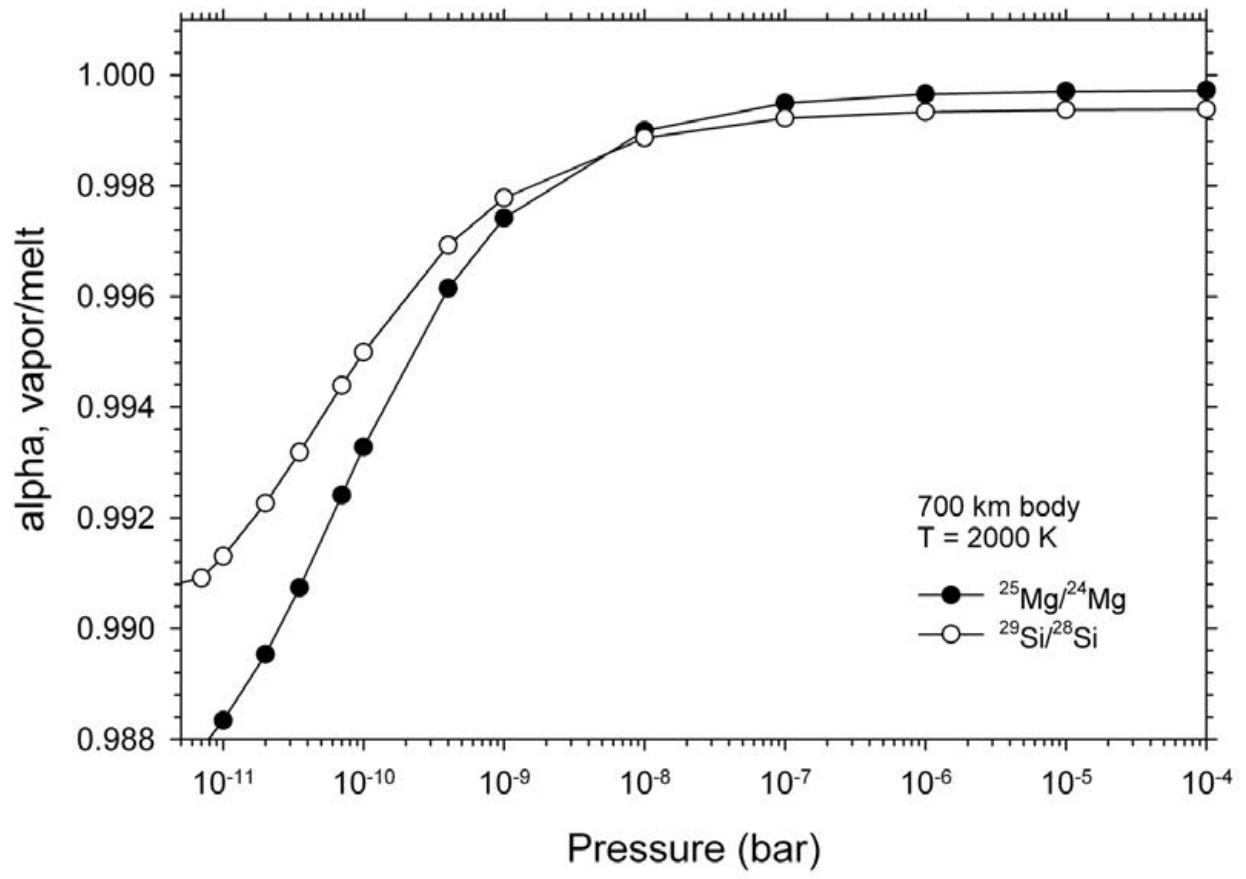

**Figure 9**. Net isotope fractionation factors for $^{25}Mg/^{24}Mg$ and $^{29}Si/^{28}Si$ as a function of total far-field gas pressure based on Equation (9). These results are for the same body depicted in Figure 8.

The scenario described above illustrates a general point – evaporation from magma oceans exposed to space on bodies large enough to retain rock vapor will produce rock vapor atmospheres. These atmospheres, while thin, are likely to be dense enough to result in approaches to melt/vapor isotopic equilibrium. However, they will generally fall short of imposing perfect equilibrium because the combination of the negative feedback on evaporation with pressure and the positive feedback with pressure on gas escape means that the pressure of the rock vapor cannot increase unabated (Figure 8). This conclusion is broadly consistent with the recent suggestions by Hin et al. (2017) and Norris and Wood (2017) that melt-vapor equilibrium explains heavy isotope enrichment in Si and Mg and the concentrations of moderately volatile chalcophile elements in bulk silicate Earth.

The approximate timescale for establishing the rock vapor atmosphere around an evaporating planetesimal is rapid. This can be illustrated using the ½ $M_{\mathrm{Pluto}}$ body described above. From the surface to the Bondi radius there are approximately 1 ×$10^{11}$ moles of SiO gas at steady state. The net evaporation rate of SiO at steady state is ~ 1×$10^9$ moles $s^{-1}$, suggesting a timescale for atmosphere formation of ~ 1 ×$10^{11}$ moles/1×$10^9$ moles $s^{-1}$ ~ 100 seconds.

## 5. Focussing on the Residual Melt Composition

The net evaporative fluxes and isotope fractionation factors described above can be combined with a physical model for evaporation in order to simulate the isotopic consequences of planetesimal magma oceans exposed to space. For this purpose we make use of the "shrinking core" model in which mass balance at an evaporating and retreating melt surface is maintained (Young et al. 1998). In this model, reprised here with slight modifications to the evaporative fluxes in order to accommodate the build up of rock vapor atmospheres, concentrations in the sphere of evaporating melt are obtained from the diffusion equation

$$\frac{\partial c_{i,\text{melt}}}{\partial t} = D_{i,\text{melt}}\left(\frac{\partial^2 c_{i,\text{melt}}}{\partial r^2} + \frac{2}{r}\frac{\partial c_{i,\text{melt}}}{\partial r}\right), \quad 0 < r < s(t) \;, \tag{18}$$

where $c_{i,\text{melt}}$ is the concentration of species $i$ in the melt at position $r$ and time $t$, $r$ refers to the radial position within the spherical body, $s(t)$ is the time-dependent radius of the body due to the retreating evaporation surface, and $D_{i,\text{melt}}$ is the diffusivity of species $i$ in the melt phase. At the surface of the evaporating sphere, mass balance requires that

$$J_{i,\text{net}} = J_{i,\text{diff}} - J_{i,\text{ablt}} \;, \tag{19}$$

where $J_{i,\text{net}}$ is the net evaporation flux for species $i$, $J_{i,\text{diff}}$ is the diffusive flux of species $i$ from the surface of the evaporating body into the interior, and $J_{i,\text{ablt}}$ is the ablative flux associated with the motion of the retreating surface due to evaporation (Figure 10). The fluxes can be written in terms of the rate of surface retreat, $\dot{s}$, such that

$$J_{i,\text{net}} = -c_{i,\text{gas}}\,\dot{s}, \; J_{i,\text{ablt}} = c_{i,\text{s}}\,\dot{s}, \text{ and } J_{i,\text{diff}} = -D_{i,\text{melt}}\frac{\partial c_{i,\text{s}}}{\partial r} \;, \tag{20}$$

where $c_{i,\text{gas}}$ and $c_{i,\text{s}}$ are the concentrations of species $i$ in the gas immediately above the melt surface and in the melt at the melt surface, respectively, and again $r$ refers to the radial distance from the center of the spherical body. The boundary condition at the evaporation surface from Equations (19) and (20) is (Young et al. 1998)

$$D_{i,\text{melt}}\frac{\partial c_{i,\text{s}}}{\partial r}\Big|_{r=s(t)} = c_{i,\text{s}}\left(\frac{J_{i,\text{net}}}{J_{i,\text{congruent}}} - 1\right)\dot{s} \;, \tag{21}$$

where $J_{i,\text{congruent}} = x_i\,\dot{s}\,/\,\hat{V}_{\text{melt}}$ is the congruent molar flux of species $i$, $x_i$ is the mole fraction of the appropriate oxide of $i$ in the melt (e.g., MgO, $SiO_2$, CaO, and $Al_2O_3$), and $\hat{V}_{\text{melt}}$ is the molar volume of the melt. The congruent flux is the flux that would obtain if there were no selective partitioning between vapor and melt whereas $J_{i,\text{net}}$ is the flux prescribed by activities in the melt and the return fluxes (Equation 7). Young et al. (1998) showed that for simultaneous evaporative mass loss at the surface and diffusive mixing in the interior, subject to the boundary condition in Equation (21), concentrations in the melt as functions of time are obtained by solving the dimensionless form of Equation (18) that includes the moving evaporating surface:

$$\tilde{s}^2\frac{\partial u}{\partial \tilde{t}} = \frac{\partial^2 u}{\partial R^2} - \beta R \tilde{s}\frac{\partial u}{\partial R}, 0 < R < 1 \;, \tag{22}$$

where $\tilde{t} = t\,D_{i,\text{melt}}\,/\,s_{\text{o}}^2$ is dimensionless time (analogous to $\xi$ diffusivity in the the melt rather than in gas), $s_{\text{o}}$ is the initial radius of the spherical body, $\tilde{r} = r\,/\,s_{\text{o}}$ is the dimensionless radial position in the melt relative to the initial radius, $\tilde{s} = s\,/\,s_{\text{o}}$, $R = \tilde{r}\,/\,\tilde{s}$ is the transformation that accommodates the moving boundary (Young et al. 1998), $\tilde{c}_{i,\,\text{melt}} = c_{i,\,\text{melt}} / c^{\text{o}}_{i,\,\text{melt}}$ is the dimensionless concentration of species $i$ in the melt relative to the initial concentration, $u = \tilde{r}\,\tilde{c}$ , and $\beta = -s_{\text{o}}\,\dot{s} / D_{i,\,\text{melt}}$ is the dimensionless rate of surface migration. $\beta$ is also the Péclet number representing the ratio of ablative to diffusive transport; $\beta >> 1$ implies mixing in the condensed phase is limited by diffusion, while $\beta < 1$ implies that the condensed phase remains well mixed during evaporation. The surface boundary condition in dimensionless form becomes

$$\frac{\partial u}{\partial R}\bigg|_{R=1} = u\left(1 - \beta\,\tilde{s}\left(\frac{J_{i,\,\text{net}}}{J_{i,\,\text{congruent}}} - 1\right)\right) . \tag{23}$$

In the present application we use reference evaporative fluxes derived from CMAS melts (Grossman et al. 2000; Richter et al. 2002; Shahar and Young 2007) to simulate evaporation of the silicate portions of planetesimals, ignoring for this work the effects of metal core formation. The reference fluxes for Mg and SiO are specified at specific bulk compositions, $P_{H2}$, and temperatures together with an activiation energy of 650 kJ/mole (Richter et al. 2002). Reference fluxes for Ca, Al, AlO and AlOH are taken from Grossman et al. (2000). Evaporation fluxes are obtained from the reference fluxes by correcting for differences in melt composition (e.g., Equation 10), temperature, and $P_{H2}$ (Shahar and Young 2007 and references therein). We ignore core formation in the present work mainly because of uncertainties in the timing of core formation relative to evaporation, saving this problem for a later installment. The rate of surface retreat due to evaporation is obtained from the total evaporative flux that is the sum of fluxes for $^{24}Mg$, $^{25}Mg$, $^{28}SiO$, $^{29}SiO$, Ca, Al, AlO, and AlOH recast as the total moles of oxides lost from the melt (e.g., total moles of MgO lost = moles $^{24}Mg$ + $^{25}Mg$ evaporated, moles of $Al_2O_3$ lost = evaporated moles (Al + AlO +AlOH)/2) , and the molar volume of the melt, $\hat{V}_{\text{melt}}$ :

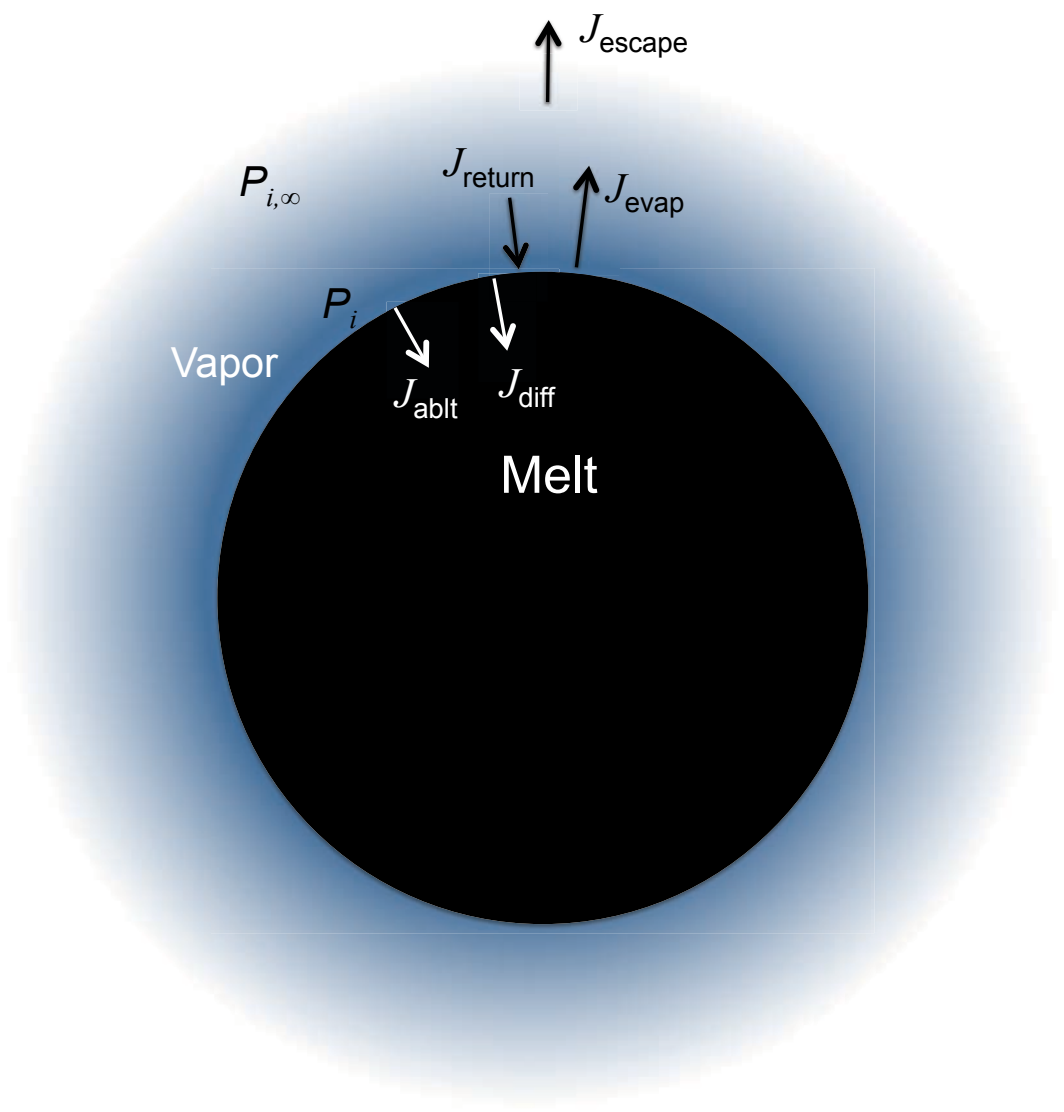

**Figure 10**. Schematic illustrating the fluxes involved in evaporation of a magma ocean body with a rock vapor atmosphere. See text for definitions.

$$\dot{s} = \frac{ds}{dt} = \hat{V}_{\text{melt}}\,J_{\text{total}} \tag{24}$$

where the molar volume is calculated using the formulation of Courtial and Dingwell (1999). Numerical solutions to Equation (22) subject to boundary condition (23) are obtained for $^{24}MgO$, $^{25}MgO$, $^{28}SiO_2$, $^{29}SiO_2$, CaO, and $Al_2O_3$ in the melt as functions of time. Evaporative fluxes are updated at regular intervals between time steps in order to capture the changes ensuing from the evolving melt compositions. In order to simulate the extensive mixing in the melt due to vigorous convection, we multiply diffusivities in the melt by the ratio of convective to conductive heat transport embodied by the Nusselt number prescribed by the Rayleigh number (e.g., $Nu = 10^9$ for the ½ $M_{Pluto}$ body, see above). This effectively eliminates any diffusive limitations to mixing of the melt during evaporation; $\beta$ is < 1 in all simulations shown here. The utility of the model described herein where the molten body is well mixed is in the calculation of the incongruent evaporation fluxes. The boundary condition in Equation (23) becomes increasingly important only where mixing is limited.

## 6. Results and Comparisons with Planetary Isotopic Compositions

In this section we apply the physical evaporation model described in §3-5 to selected combinations of planetesimal size and far-field ambient pressures. The results can be compared with the isotopic compositions of various differentiated bodies in the solar system relative to CI chondrites and E chondrites (e.g., Figure 1). The goal is to narrow down the plausible evaporation conditions that might have played a role in establishing the chemical and isotopic composition of Earth and other differentiated bodies. These calculations invoke the relationships between planetesimal size and rock vapor atmospheres developed in §3 and §4. We confine our calculations to the CMAS system for which evaporation data are well calibrated for now. The relationships between Mg and Si isotopes are likely to be well represented by the results in the CMAS system.

### *6.1 Isotope fractionation controlled by evaporation*

In order to model the evaporation of the silicate portions of planetesimals we considered two plausible starting compositions, CI chondrite and EL chondrite. The compositions we use (Jarosewich 1990; Newsom 1995) are projected into the CMAS chemical system by renormalizing the concentrations of MgO, $SiO_2$, CaO and $Al_2O_3$ to 100% (e.g., Table A1). The Mg and Si isotopic compositions of evaporating bodies with a CI bulk composition projected into the CMAS system are shown in Figure 11. The isotopic compositions of bulk silicate Earth, Mars, HED meteorites (Vesta), and the angrites from Figure 1 are shown for comparison. Four evaporation scenarios are shown that span possible combinations of planetesimal mass and pressure, including 50 km bodies (not withstanding that such small bodies may not form magma oceans) evaporating into $10^{-8}$ total pressure and $10^{-4}$ bar of $H_2$, and 700 km bodies evaporating into these same far-field pressures. By total pressure we include gas comprising collision partners but that might not be reactive, unlike $H_2$. The higher $P_{H2}$ simulations represent evaporation of the planetesimals in the presence of the protostellar disk gas.

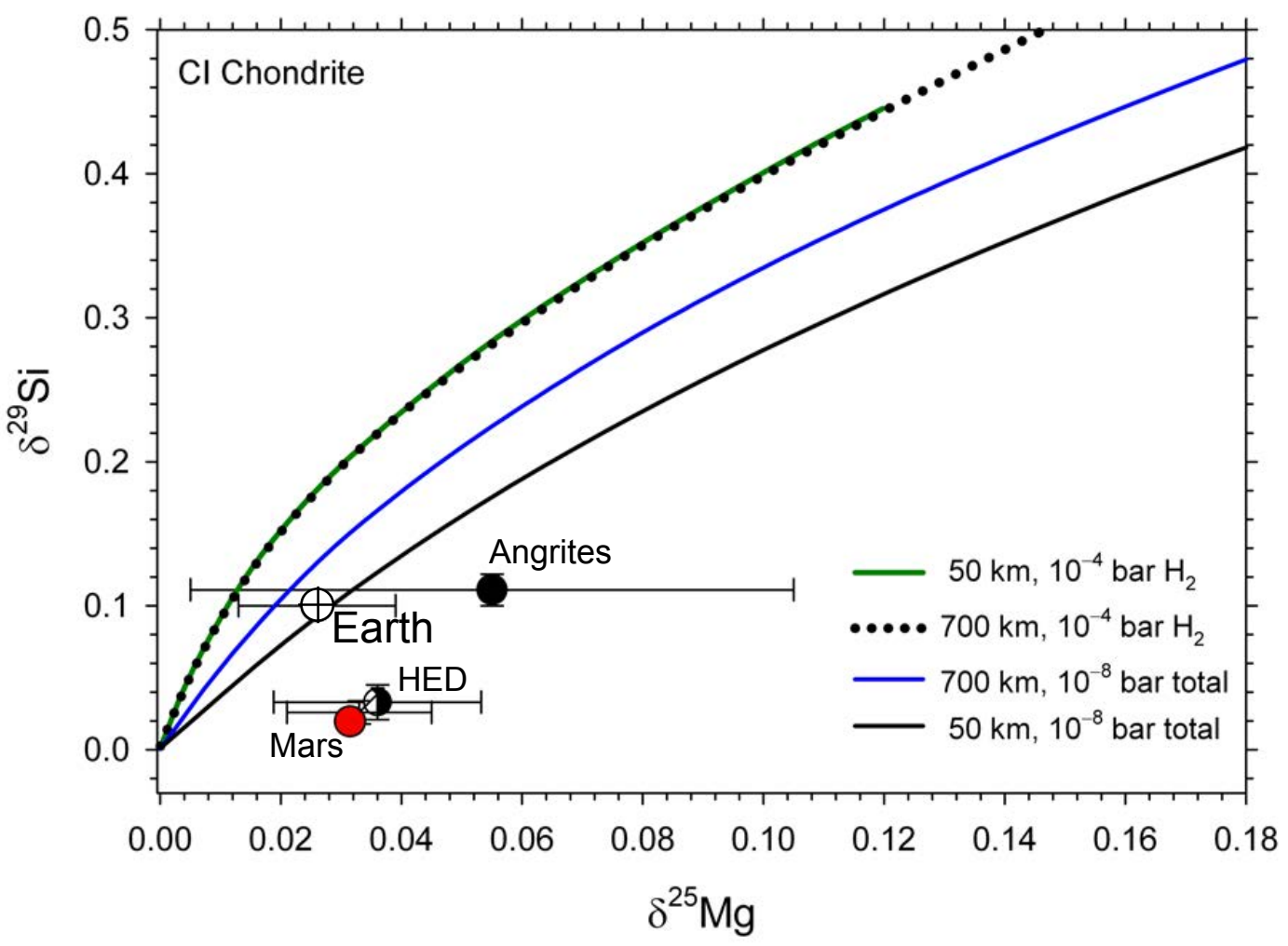

**Figure 11**. Model curves depicting the Mg and Si isotopic compositions of planetesimals of CI composition subjected to evaporation. Each point on the curves represents a particular time during the evaporation process. Four models representing two different planetesimal radii and two different pressure environments are shown. The 700 km model at $10^{-8}$ bar represents evaporation that forms a steady-state rock vapor for the ½ $M_{Pluto}$ body absent protostellar nebular gas. The 700 km and 50 km bodies at $10^{-4}$ bar of $H_2$ represent evaporation in the presence of protostellar gas. Estimates for the isotopic compositions of silicate Earth, Mars, the HED parent body (Vesta), and the angrite parent body are shown for reference (same data sources as for Figure 1).

Higher pressures of hydrogen gas affect evaporation in two opposing ways. Hydrogen pressure induces a significant return flux like any other gas, but it also increases evaporative fluxes (Young et al. 1998; Richter et al. 2002). The $10^{-8}$ total pressure simulates the steady-state rock vapor atmosphere for the 700 km body (1/2 $M_{\mathrm{Pluto}}$) in the absence of nebular gas and a 50 km body at this same pressure is presented for comparison. All of the calculations presented here are performed with an approximate melt temperature of 2000 K. Detailed parameters associated with the models can be found in the Appendix in Table A2.

At face value all four of the model curves intersect the bulk silicate Earth within uncertainties. None of the curves intersect the compositions of Mars, Vesta, or the angrites. It is likely, however, that the Mg isotopic compositions used for these bodies are affected by fractional crystallization or partial melting. The analyzed samples of Mars, Vesta, and angrites are all basalts. The data of Hin et al. (2017) suggest that terrestrial basalts tend to be higher in $\delta^{25}$Mg than terrestrial ultramafic rocks by of order ~ 0.01 ‰. This is not unexpected because equilibrium fractionation in $^{25}$Mg/$^{24}$Mg between forsterite and pyroxene is on the order of 0.02 ‰, with forsterite having the lower $\delta^{25}$Mg values (Schauble 2011). Similarly, in mantle rocks, olivine has lower $\delta^{25}$Mg values than both whole rocks and coexisting pyroxene by a few tens of ppm (Young et al. 2009). These observations suggest that fractional crystallization of olivine, or partial melting leaving behind an olivine residue, could lead to $\delta^{25}$Mg values greater than the source mantle by as much as ~ 0.02 ‰. For this reason we will focus on matching the isotopic and chemical composition of bulk silicate Earth (BSE) and assume for now that the basalts representing the other bodies are affected by Mg isotope fractionation relative to the bulk parent bodies.

Since all four evaporation scenarios depicted in Figure 11 pass through the error bars for BSE, another discriminator is required to evaluate the relative viabilities of the different evaporation conditions. This discriminator can come from the relationships between $\delta^{25}$Mg and $F_{\mathrm{Mg}}$ and between $\delta^{29}$Si and $F_{\mathrm{Si}}$. The models for these relationships are compared with BSE in Figure 12. In order to obtain $F_{\mathrm{Mg}}$ and $F_{\mathrm{Si}}$ for BSE we rely on the refractory nature of Ca. Calcium in effect does not evaporate except under the most extreme depletions of other elements (e.g., $J_{\mathrm{Ca,net}}$ /$J_{\mathrm{Ca,congruent}}$ typically is ~ $10^{-4}$, Table A2). Therefore, to a good approximation $F_{\mathrm{Mg}} = n_{\mathrm{Mg}}/(n_{\mathrm{Mg}})_{\mathrm{o}}$ = ([Mg]/[Ca])/([Mg]/[Ca])$_{\mathrm{o}}$, where the brackets represent concentrations and the subscript zero signifies the initial condition. An analogous expression applies to Si. Using this Ca normalization, BSE has $F_{\mathrm{Mg}} = 0.88$ and $F_{\mathrm{Si}} = 0.86$ relative to a CI chondrite initial composition, implying loss of 12 % Mg and 14% Si from the planetesimals accreted to form Earth.

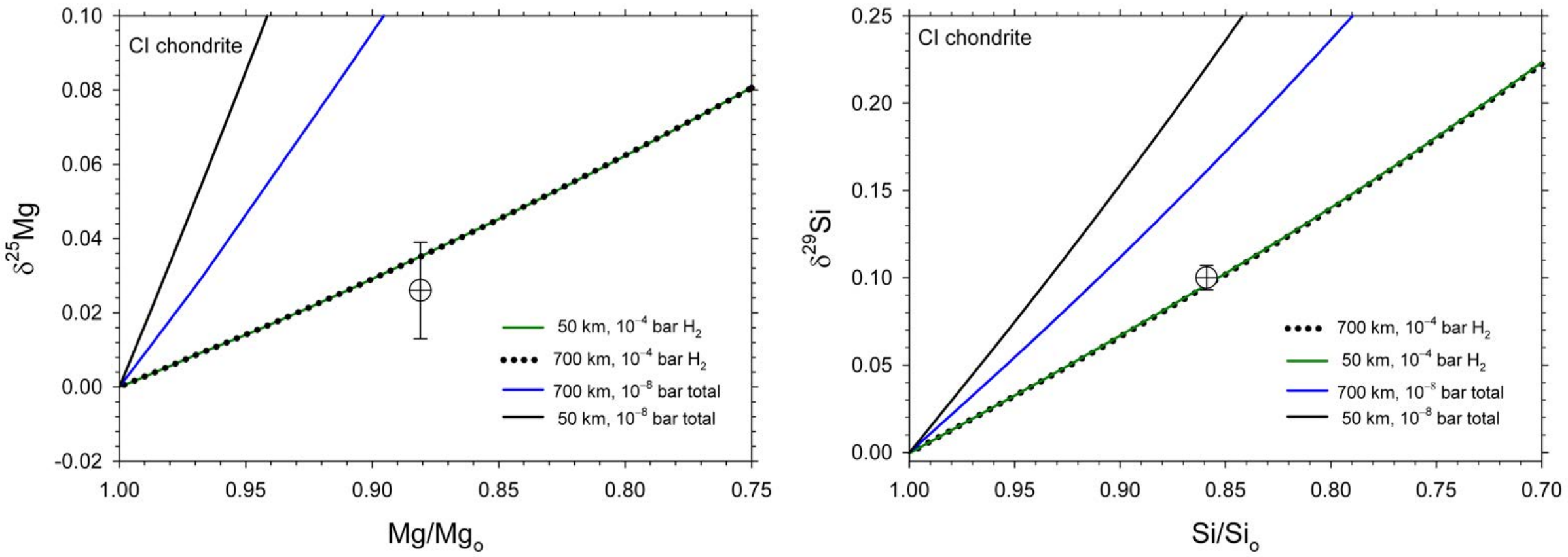

**Figure 12**. Mg and Si isotopic shifts with evaporation of CI-like planetesimals vs. fractions of Mg and Si remaining in the melt, respectively. The composition of Earth as represented by BSE relative to a CI chondrite starting material is shown for comparison (circle with cross).

Results summarized in Figure 12 show that only the high-pressure evaporation models fit the isotopic and mass-loss characteristics of the bulk silicate Earth. The results imply that evaporation of planetesimal melts in the presence of protostellar nebular $H_2$ is required to explain the Mg and Si isotopic and elemental concentrations of bulk Earth relative to CI chondrites. The similarity between the 700 km (1/2 $M_{\mathrm{Pluto}}$) and 50 km-sized bodies at high $P_{\mathrm{H2}}$ values of $10^{-4}$ bar results from the fact that the net evaporation fractionation factors for both body sizes closely approximate $1/\alpha_{\mathrm{eq}}$, i.e., the equilibrium vapor/melt fractionation factor. This is because at $10^{-4}$ bar, the mean free path □ for the gas molecules at relevant temperatures is on the order of $10^{-3}$ meters. Since the evaporation factors $\gamma_{□}$ for Mg and Si have values of about 0.02 to 0.2 (Richter et al. 2002; Fedkin et al. 2006), the critical planetesimal radius for 50% saturation, ~ $\lambda/\gamma_{□}$, is < 1 meter. For the 700 km body, the difference in isotope fractionation vs. mass loss relationships at $10^{-8}$ bar and at $10^{-4}$ bar is the difference between ~ 95 % saturation and ~ 99% saturation, respectively.

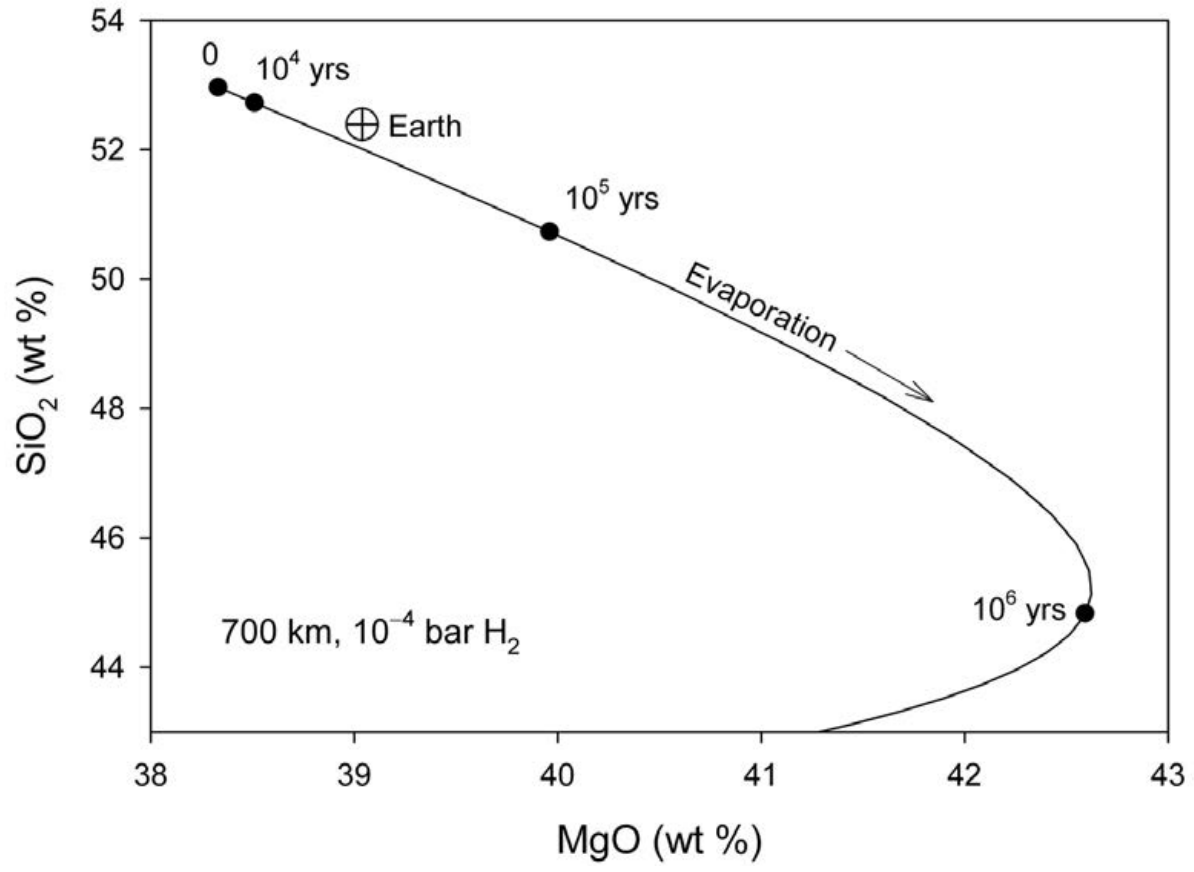

**Figure 13**. Plot of weight per cent $SiO_2$ vs MgO for the ½ $M_{\mathrm{Pluto}}$ evaporation model based on a CI starting composition projected into the CMAS chemical system. Earth is shown for comparison (circle with cross). Black dots mark the time evolution, suggesting a cessation of evaporation between ~ $10^4$ and $10^5$ years.

The duration of evaporation required to achieve the isotopic and chemical shifts for Mg and Si evidenced by the bulk silicate Earth is about $10^4$ years (Figure 13). The evaporation timescales are relatively insensitive to the radii of the bodies when they evaporate into $10^{-4}$ bar $H_2$ gas.

Evaporation models like those shown in Figures 11 through 13 were also calculated for an EL starting material. Results show that a match with the bulk isotopic and chemical composition of Earth was not possible using EL chondrite as the starting material. Model results are broadly similar to those for the CI starting composition. However, because the E chondrites have higher concentrations of $SiO_2$ compared with the CI chondrites, and between 2 and 4 times lower CaO concentrations than CI chondrites (depending upon whether EL or EH bulk compositions are used), the calculated mass fractions of Mg and Si remaining for the Earth are both < 0.5 and cannot be matched by the models. The problem is mitigated somewhat if one uses Al as the normalizing refractory element, resulting in $F_{\mathrm{Si}}$ for Earth of 0.6. However, this is still too much mass loss for the degree of isotope fractionation. The results for this case showing the mismatch between the E-chondrite models and the bulk silicate Earth relative to E chondrite are shown in Figure 14. We conclude that evaporation of planetesimals with E-chondrite-like compositions cannot alone account for the Mg and Si isotopic composition of the Earth and the apparent mass losses of these elements.

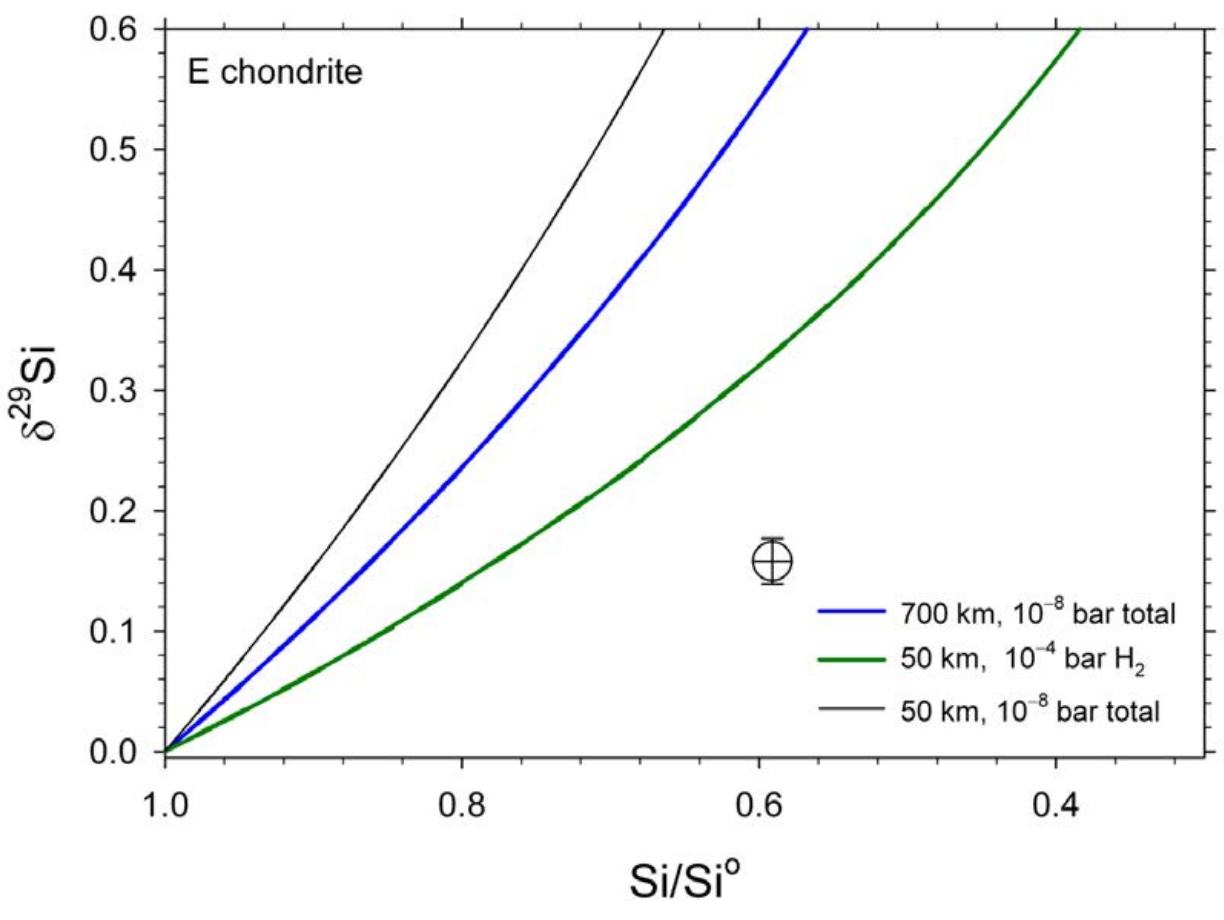

**Figure 14**. Mg and Si isotopic shifts with evaporation of E chondrite-like planetesimals vs. fractions of Mg and Si remaining in the melt, respectively. The composition of Earth relative to the EL chondrite starting material is shown for comparison (circle with cross).

*6.2 Isotope fractionation controlled by atmospheric loss*

The presence of a steady-state rock-vapor atmosphere can have additional isotopic consequences for the melt in principle. In the fiducial ½ $M_{\mathrm{Pluto}}$ case used for illustration throughout this paper, escape velocities are attained at an altitude within the mixing layer (troposphere) of the rock vapor atmosphere (Figure 6). Isotope fractionation caused by hydrodynamic escape could therefore be transmitted downward to the melt-vapor interface by convection. Exchange of isotopes at this interface is expected because the rate of return flux from gas to melt nearly matches the evaporative flux from the melt at the steady-state atmospheric pressure, as described above. The result is a mechanism for melt-vapor exchange that operates on a timescale that is faster than evaporation alone (Figure 15). An analysis of the potential isotopic effects of this process is given here. We find that the effects for hydrodynamic escape are negligible but could be significant for Jeans' escape.

In the context of our ½ $M_{\mathrm{Pluto}}$ scenario where the only gas surrounding the body is that produced by evaporation of the magma ocean, the number of molecules of species $i$ in the atmosphere, $N_{i,\mathrm{gas}}$, is controlled by the net flux from the melt surface balanced against the rate of loss by hydrodynamic escape, leading to

$$\frac{dN_{i,\,\mathrm{gas}}}{dt} = 4\pi s^2 J_{i,\,\mathrm{net}} - 4\pi r_{\mathrm{B}}^2 C_{\mathrm{S}}(r_{\mathrm{B}}) n_{i,\,\mathrm{gas}} \quad , \tag{25}$$

where $n_{i,\mathrm{gas}}$ is the number density for species $i$ in the gas. Because the rock vapor builds up rapidly, we are concerned with the steady-state where

$$n_{i,\,\mathrm{gas}}(t=\infty)=\frac{J_{i,\,\mathrm{net}}}{C_{\mathrm{S}}(r_{\mathrm{B}})}\left(\frac{s}{r_{\mathrm{B}}}\right)^{2} . \tag{26}$$

The same mass balance equations apply for the heavy isotopic species. Denoting the heavy isotopic species with the prime superscript, the steady-state isotope ratio in the atmosphere is

$$\left(\frac{n'_{i,\,\mathrm{gas}}}{n_{i,\,\mathrm{gas}}}\right)_{t=\infty}=\frac{J'_{i,\,\mathrm{net}}}{J_{i,\,\mathrm{net}}}\frac{C_{S}(r_{B})}{C'_{S}(r_{\mathrm{B}})} . \tag{27}$$

The right-hand side of Equation (27) is the net evaporation isotope fractionation factor divided by the hydrodynamic escape fractionation factor. From the evaporation calculations for our steady-state atmosphere model we obtain $J'_{i,\,\mathrm{net}}/J_{i,\,\mathrm{net}}=0.9991$, representing the $^{25}$Mg/$^{24}$Mg ratio leaving the melt surface; there is a slight bias against evaporating $^{25}$Mg relative to $^{24}$Mg at near-equilibrium conditions (Table A1). By itself, this effect would deplete the rock vapor in $^{25}$Mg relative to $^{24}$Mg.

Hydrodynamic escape involves a wholesale motion of the molecules as they escape with the result that fractionation relative to the average molecular mass depends on the ratio of the mutual diffusivity to the magnitude of the flux (Hunten 1973; Zahnle and Kasting 1986; Hunten et al. 1987). Because of the drag forces among the molecules, the sound velocities for the different species in Equation (27) are more similar to one another than their dependence on molecular mass might otherwise suggest. Using the analysis from Zahnle and Kasting (1986, see also Hunten 1973), the fractionation factor, or ratio of escape rates, between a species with molecular mass $m_i$ and the average molecular mass is

$$\alpha_{hydro}\sim 1-\frac{GM(m_i-m_{\mathrm{Avg}})D_i n_{\mathrm{Gas}}}{(C_{\mathrm{S}}(r_{\mathrm{B}})n_{\mathrm{Gas}})r_B^{\,2}k_{\mathrm{b}}T} , \tag{28}$$

where the form of the hydrodynamic escape flux is preserved in the denominator, $n_{\mathrm{Gas}}$ refers to the number density of the gas in total, and we employ the high-flux approximation used by Zahnle and Kasting for this application. With $m_{\mathrm{Avg}}$ = 0.034 kg/mole for the rock-vapor atmosphere, and applying Equation (28) for $^{25}$Mg and $^{24}$Mg, the $^{25}$Mg/$^{24}$Mg fractionation factor is 0.9999939, or −0.006 ‰. Similar results obtain for $^{29}$SiO/$^{28}$SiO and for different versions of Equation (28) based on different assumptions (e.g., sub-equal mixing ratios). Substitution of this fractionation factor into Equation (27) shows immediately that hydrodynamic escape should not impart measurable isotope fractionation effects for a rock-vapor atmosphere (Figure 15).

In the case of Jeans' escape, fractionation is much larger. Following the Jeans' escape analog to Equation (25), the steady-state atmospheric isotope ratio in this case is related to the evaporation rates and the Jeans' escape velocities such that

$$\left(\frac{n'_{i,\,\mathrm{gas}}}{n_{i,\,\mathrm{gas}}}\right)_{t=\infty} = \frac{J'_{i,\,\mathrm{net}}}{J_{i,\,\mathrm{net}}}\frac{\omega}{\omega'} \quad . \tag{29}$$

The result for our ½ $M_{\mathrm{Pluto}}$ model is that the steady-state isotope ratio in the atmosphere, accounting for the small depletion in the heavy isotope from evaporation and the greater enrichment in the heavy isotope by Jeans' escape, is 0.9991 × 1.0466 = 1.0456. The atmosphere could therefore be enriched relative to the initial $^{25}$Mg/$^{24}$Mg of the system by 45.6 ‰ ($\delta^{25}$Mg = 45.6 ‰) if well mixed.

The rate of isotope exchange between the heavy isotope enriched vapor and the melt is nearly equal to the net evaporation rate since the return flux from the vapor to the melt is 0.95 times the evaporation flux (Table A2). For the concentrations of Mg in the melt and the vapor in our fiducial model, the fraction of Mg contained in the atmosphere relative to the total in the melt and atmosphere at steady state is $N_{\mathrm{Mg,atm}}/(N_{\mathrm{Mg,\,atm}} + N_{\mathrm{Mg,\,melt}}) = 4 \times 10^{-12}$. This value can be compared with the fraction of Mg required to explain Earth's $\delta^{25}$Mg, $\delta^{25}Mg_{\oplus}$, of 0.026 ‰ relative to chondrites, if it were due entirely to exchange between a rock-vapor atmosphere affected by Jeans' escape and its progenitor magma ocean. The fraction of exchanged Mg in this case is derived using the conservation equation for Mg isotopes:

$$\delta^{25}Mg_{\oplus} = 0.026 = x_{\mathrm{Mg,\,Ex}} 45.6 + \left(1 - x_{\mathrm{Mg,\,Ex}}\right)0 \tag{30}$$

where $x_{\mathrm{Mg,\,Ex}} = N_{\mathrm{Mg,Ex}}/(N_{\mathrm{Mg,\,Ex}} + N_{\mathrm{Mg,\,melt}})$ and $N_{\mathrm{Mg,\,Ex}}$ refers to the amount of magnesium in the melt derived from the atmosphere by isotopic exchange, and the $\delta^{25}$Mg values for the rock-vapor atmosphere and the original melt are 45.6 ‰ and 0 ‰, respectively. Solving Equation (30) for

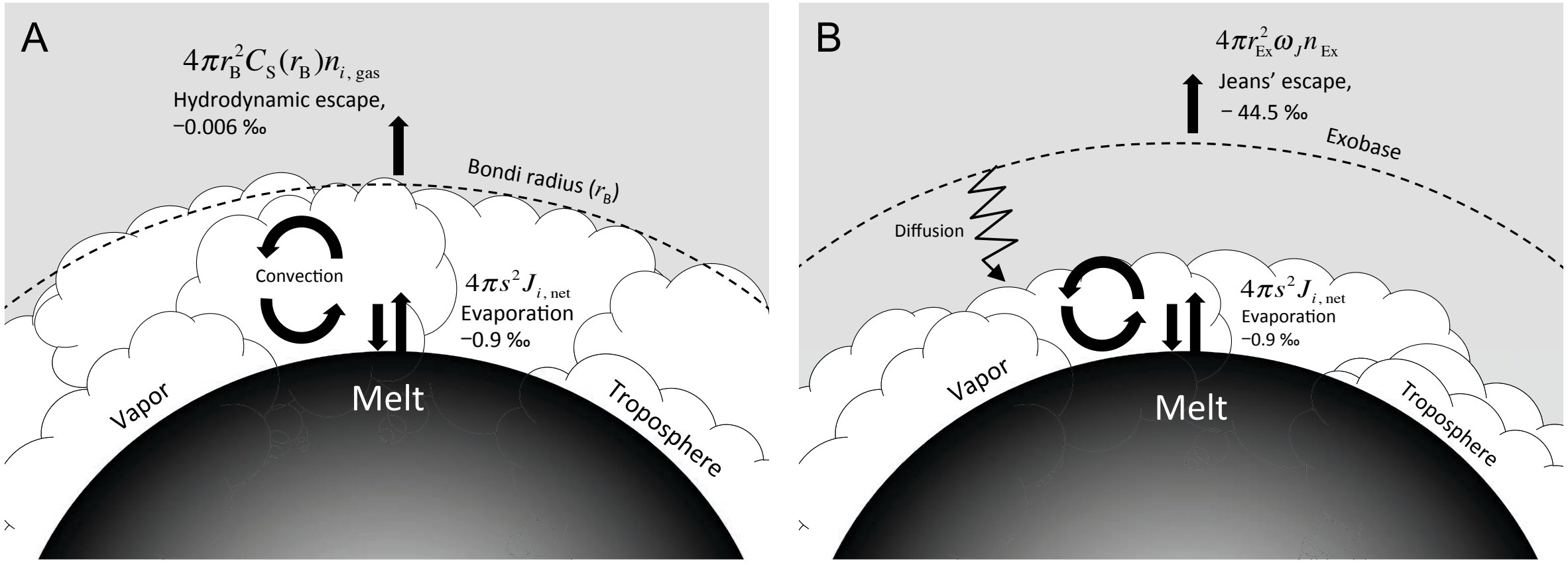

**Figure 15**. Schematic illustration of the balance between evaporative losses of Mg gas from a molten silicate surface and concurrent losses of Mg from the vapor by either hydrodynamic escape (A) or Jeans' escape (B). The $^{25}$Mg/$^{24}$Mg isotopic fractionation of the gas emanating from the melt (-0.9 ‰ relative to the melt) and that of the gas leaving the atmosphere (-0.006 and -44.5 ‰ for hydrodynamic escape and Jeans' escape relative to the gas, respectively) are shown in both cases. The positions of the Bondi radius and the exobase relative to the mixed layer (troposphere) are indicated for the two different scenarios by the dashed curves. Mixing by convection is indicated by the circular arrows and possible mixing by diffusion is shown by the zigzag arrow.

$x_{\mathrm{Mg,\,Ex}}$ yields a value of 0.0006, corresponding to 1.7 $\times 10^{19}$ moles of Mg. The disparity between the steady-state fraction of total Mg in the gas phase compared with the much larger quantity of atmospheric Mg needed to explain an increase in $\delta^{25}$Mg of ~0.02 ‰ requires continuous exchange over some time interval. The rate of vapor-melt exchange of Mg based on our near-equilibrium calculation is 2.5 $\times 10^{18}$ moles Mg/s from the ratio of Mg flux to total flux (Table A2). The required interval for exchange is therefore 1.7 $\times 10^{19}$ moles Mg/ 2.5 $\times 10^{8}$ moles/s = 6.8 $\times 10^{10}$ seconds, or 2,156 years. A similar calculation for the isotopes of Si in which the terrestrial $\delta^{29}$Si value of 0.1 ‰ relative to chondrites is attributed to vapor-melt exchange yields a similar timescale (well within uncertainties).

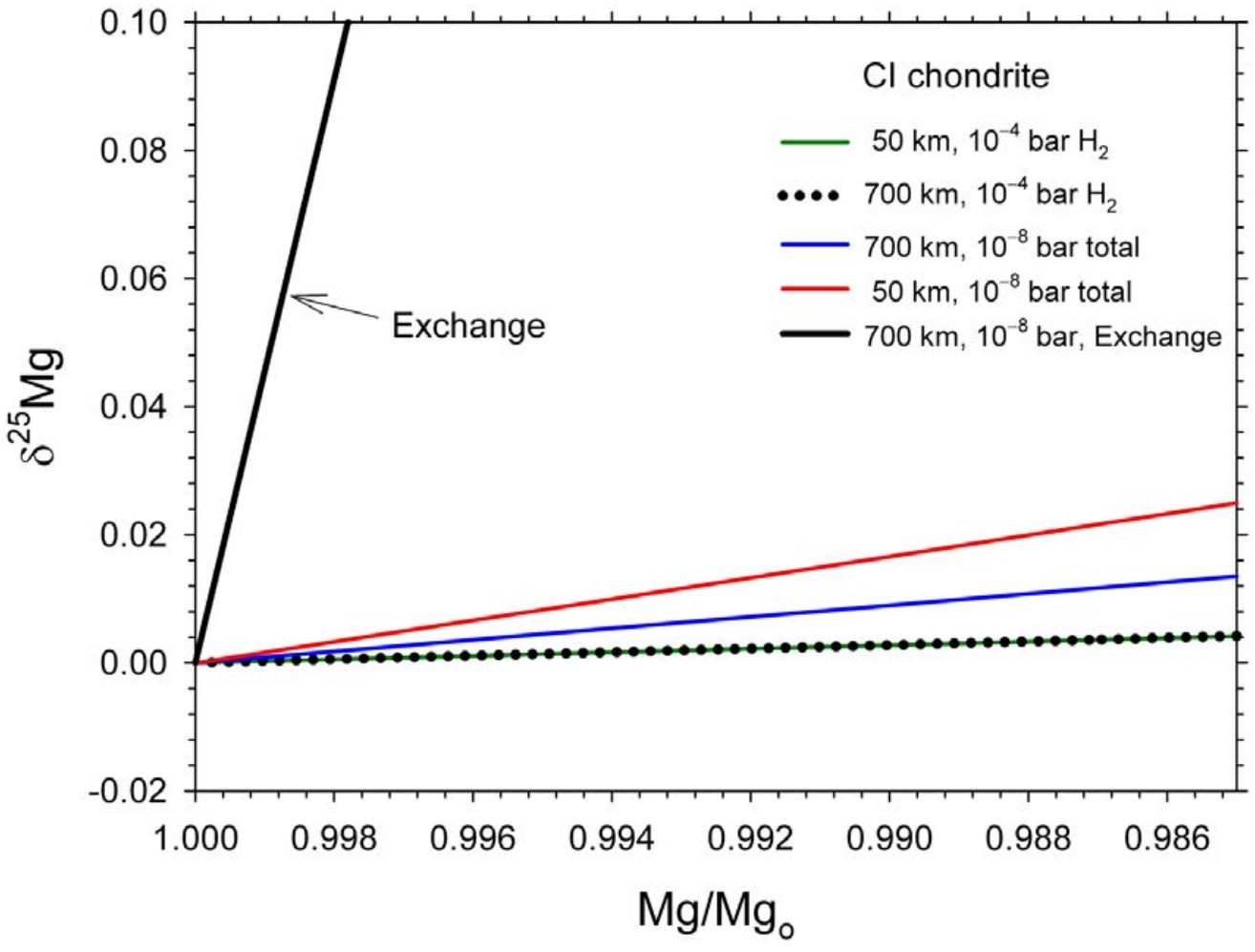

**Figure 16**. Melt $\delta^{25}$Mg vs. Mg/Mg$_o$ due to exchange between melt and a steady-state rock vapor atmosphere affected by Jeans' escape (steep, heavy black curve). Calculations are described in the text. Curves from Figure 12 are shown for comparison.

While the Mg and Si isotopic composition of Earth relative to chondrites could be explained by Jeans' escape and exchange with melt if there were sufficient mixing in the atmosphere as a whole, the apparent mass losses of Mg and Si cannot. The rate of heavy isotope enrichment of a melt due to exchange with a vapor affected by Jeans' escape is much more efficient than for net evaporation alone. Heavy isotope enrichment is accelerated by the large fractionation associated with Jeans' escape to the point where isotope exchange occurs with negligible losses in the mass of melt. In the example for Mg described above, the ratio of Mg remaining to initial Mg, Mg/Mg$_o$ corresponding to the required shift in melt $\delta^{25}$Mg of 0.026 ‰ is 0.9994, or ~0.06% loss of Mg compared with the 12% Mg loss required by evaporation alone for the Earth (Figure 16). The result is similar for Si. If atmospheric escape controlled heavy isotope enrichment in Earth's building blocks, a mechanism unrelated to isotope fractionation would be required to explain the Earth's concentrations of Mg and Si relative to Ca and Al.

Another caveat pertains to the circulation of the atmosphere in the mixing layer. The motion of gas parallel to the vapor-melt interface could decrease or even eliminate exchange across the phase boundary. It would also result in an approach to free evaporation if the velocity parallel to the melt surface was large. The likelihood for these effects could be assessed using the gas transport Péclet number, $Pe = u\Lambda/D$, where $u$ is the gas velocity in the boundary region adjacent the melt surface, $\Lambda$ is the depth of the gas boundary layer adjacent the melt, and $D_i$ is the gas-phase diffusivity for the evaporating species of interest. $Pe >> 1$ indicates advection dominates over diffusion. Under these conditions, the return flux to the melt would be inhibited, limiting exchange and driving isotope fractionation towards free evaporation. Estimating the gas-phase convective velocity and boundary layer thickness is not straightforward. The possibility that convection extending near to the melt surface disrupts gas-phase diffusive transport immediately above the melt surface requires further study.

## 7. Discussion

### *7.1 Timescale of evaporation*

The evaporative timescale of ~$10^4$ years required to explain the isotopic and elemental concentrations of Earth are well within the ~ $10^5$ year longevity of planetesimal surface magma oceans produced by $^{26}$Al decay for an accretion time of 0.2 Myr (Figure 2). The longevity of the magma oceans will have varied with accretion times. Average evaporative timescales may therefore simply be the result of the combination of typical planetesimal accretion times, the rate of heating by $^{26}$Al decay, thermal diffusivities in rocks and their melts, and thermal radiation at the surfaces of the bodies. The presence of typical protostellar gas does not alter this conclusion since the $H_2$ gas is optically thin from the Hill radii to the surfaces of the planetesimals.

The finding that evaporation models that best replicate the isotopic composition of Earth rely on the relatively high pressures that can only have come from the protostellar nebula is also consistent with the magma ocean formation by $^{26}$Al decay. Accretion times of < 1 Myr required for magma oceans generated by $^{26}$Al decay in turn require that these bodies accreted in the presence of the protostellar nebular gas because the e-fold time for disk longevity is about 4 to 5 Myr (Pecaut and Mamajek 2016).

### *7.2 Isotopic effects of core formation*

The isotopic effects of core formation need to be taken into account both for the differentiated bodies that accreted to form the Earth and the differentiation of the Earth itself. In both cases, core-formation effects should be added to the current isotopic composition of the bulk planet in order to calculate the extent of isotope fractionation during evaporation. In the case of silicon, experimental data show that there is a large isotope fractionation between metal and silicate that could affect the bulk silicate Earth $^{29}$Si/$^{28}$Si (Shahar et al. 2009; Shahar et al. 2011). The extent of isotope fractionation that would be expressed in Earth's silicates is a function of the concentration of Si in the core that is in turn a function of oxygen fugacity ($fO_2$), temperature, and the pressure during core formation (Ziegler et al. 2010). For example, core formation may have involved equilibration between rock and metal at a temperature of ~ 3000 K, a pressure of ~ 40 GPa, and an $fO_2$ that is 1 log($fO_2$) unit below the Iron-Wüstite (Fe-FeO) $fO_2$-$T$ buffer ($\Delta$IW = −1 using non-ideal activities) based on silicate-metal partitioning of some siderophile elements (Wade and Wood 2005). Under these conditions the calculations summarized in Ziegler et al. (2010) suggest 3 wt. % Si in the core and a $\delta^{29}$Si for the BSE of 0.03 ‰ relative to the bulk Earth (i.e., chondrite). This isotopic shift of +0.03 ‰ per amu should be subtracted from the current silicon isotopic ratios of the bulk silicate Earth to correct for the fractionation during core formation, although it is small in comparison to the larger differences between BSE and chondrites.

A value of 3 wt. % Si in the core is not enough to explain the difference between BSE and chondrite Mg/Si ratios and would suggest that the bulk Earth Mg/Si is also not chondritic. The bulk silicate Earth has an atomic Mg/Si ratio of 1.25 (McDonough and Sun 1995) compared with the CI chondrite ratio of 1.04 (Lodders 2003), implying that some Si is missing from the mantle. Using this value for the BSE, the bulk Earth would have a non-chondritic atomic Mg/Si ratio of 1.18 if there were just 3 wt. % Si in the core. For the Earth to have both a CI chondrite bulk atomic Mg/Si ratio of 1.04 and the observed non-chondritic BSE Mg/Si ratio of 1.25, the core would have to be harboring 9 wt. % Si. Such high Si concentrations have been suggested previously (e.g., Rubie et al. 2011). These higher concentrations of Si would have led to a shift in BSE $\delta^{29}$Si relative to chondrites that is consistent with observations, and would imply $\Delta$IW ~ −1.5 rather than −1 at 40 GPa (i.e., more reducing conditions during core formation, figures 4 and 5 from Ziegler et al. 2010). Conversely, the lower value of 3 wt. % Si in the core is supported by another recent estimate involving core formation under relatively oxidizing conditions resulting in ~ 4 wt. % O and 3 wt. % Si in the core (Badro et al. 2015).

Magnesium in the core is not a likely explanation for the non-chondritic BSE Mg/Si ratio. Magnesium enters the metal at very high temperatures and is expected to be present in only trace concentrations at ca. 3000 K (Wahl and Militzer 2015; O'Rourke and Stevenson 2016).

From this discussion we conclude that if the lower estimates for the Si content of the core are correct, then the bulk Earth, and not just the BSE, may be missing Si relative to Mg and chondrites (the problem is only exacerbated by comparing BSE to E chondrites rather than CI chondrites). The BSE silicon isotope ratio is therefore likely the result of both core formation and a process related to accretion, and that process may have been either evaporative loss of Si or exchange between a rock vapor atmosphere and melt in Earth's antecedent planetesimals.

*7.3 Implications for moderately volatile elements*

The fractional losses of Mg and Si from Earth's building blocks implied by the terrestrial isotopic compositions of Mg and Si have collatoral implications for the more volatile elements. Here we evaluate those implications for potassium as a representative of the moderately volatile elements. In the absence of a detailed calculation for K, we can apply a simple scaling to the $^{25}$Mg/$^{24}$Mg net fractionation factor to estimate the compatible $^{41}$K/$^{39}$K net fractionation factor where $\ln(\alpha_{\square\square\square}{}^{41}\mathrm{K}/^{39}\mathrm{K})/\ln(\alpha_{\square\square\square}{}^{25}\mathrm{Mg}/^{24}\mathrm{Mg}) = m^{25}\mathrm{Mg} \times m^{26}\mathrm{Mg} \times (m^{41}\mathrm{K} - m^{39}\mathrm{K})/(m^{39}\mathrm{K} \times m^{41}\mathrm{K} \times (m^{25}\mathrm{Mg} - m^{24}\mathrm{Mg}))$ and where $m$ is the mass of the indicated species. While this scaling does not work for comparing fractionation factors of Mg and SiO because it ignores the zero-point energy effects for SiO, it is likely to be a reasonable approximation in so far as both Mg and K volatilize as the atomic species. For our steady-state rock vapor model for the ½ $M_{\mathrm{Pluto}}$ body, $\alpha_{\square\square\square}{}^{25}\mathrm{Mg}/^{24}\mathrm{Mg} = 0.99972$, yielding $\alpha_{\square\square\square}{}^{41}\mathrm{K}/^{39}\mathrm{K} \sim 0.99979$. We can apply a Rayleigh distillation model in this instance if we can estimate the fraction of K remaining. The bulk Earth K/Al relative to that of CI chondrite suggests an $F$ value for K of ~ 0.2. With the estimated fractionation factor and $F$ value, Rayleigh fractionation suggests a shift in $\delta^{41}$K relative to the initial isotopic composition of the planetesimals of 0.34 ‰ due to evaporation. This is approximately three times greater than the most recent data would suggest for the difference between Earth and chondrites (Wang and Jacobsen 2016). This rough calculation suggests that Earth contains a fraction of K that did not experience the level of evaporative loss expected on the basis of the apparent losses of the more refractory Mg and Si. Hin et al. (2017) arrived at this same conclusion.

The vapor pressure of K should be about 100 times that of Mg at the temperatures of interest (Fedkin et al. 2006), suggesting that the loss of K relative to Mg among planetesimal progenitors of the Earth was more extreme than the present-day concentrations of K and Al would imply. We verified this conclusion by simulating the evaporation of K by scaling the evaporation rate of K to that of the other elements based on equilibrium vapor pressures as described in §5. For the timescale of ~ 2 × 10$^4$ years indicated by the results for Mg and Si (Figure 13), the ratio of K concentration in the evaporated body relative to the initial concentration, $[K]/[K]_o$, is ~ $2 \times 10^{-3}$. For practical purposes, this permits the first-order approximation that all terrestrial K was lost during evaporation and later returned by unevaporated material, providing a constraint on the amount of material added to the proto-Earth that was not affected by evaporation. CV chondrites and various ordinary chondrites provide useful endmembers for unevaporated material in that they possess regular patterns of elemental concentrations that correlate well with 50% condensation temperatures. The fraction of Earth composed of unevaporated material is then $\sim [K]_{\oplus} / [K]_{\mathrm{Chond}}$ ~ 0.25 based on published concentrations for chondrites and bulk Earth. Mass balance that includes the fraction of unevaporated material as a diluent requires that the $\delta^{25}$Mg of the evaporated material should have been 0.026 ‰ rather than the present-day terrestrial value of 0.020 ‰. This difference is small in that the modestly higher $^{25}$Mg/$^{24}$Mg value for evaporated material implied by the need to add unevaporated material is within the error of the Mg isotope ratio for bulk Earth.

*7.4 Sources of decoupling isotope fractionation and mass loss*

The coupling of isotope fractionation and mass loss as depicted in Figures 12 and 14 reflects isotope fractionation at the melt-gas interface at temperatures between the solidus and liquidus of chondritic liquids. The degrees of fractionation associated with mass losses presented here differ from those reported by Hin et al. (2017). Those authors were investigating the cumulative effects of multiple collisions. Collisions can lead to a relative "decoupling" of mass loss of moderately refractory elements from isotope fractionation because of the prospects for very high temperatures well above those of the melt liquidus that greatly suppress isotope fractionation. If equilibrium melt-vapor isotope fractionation is assumed, and temperatures are as high as 3500 K, as in the Hin et al. study, isotope fractionation associated with mass loss is reduced considerably. This will lead to an effective decoupling between mass fractionation and elemental abundances. Our focus on the consequences of internal heating where temperatures generated by $^{26}$Al decay are between ~ 2000 K and ~1500 K does not permit this circumstance. Super-liquidus temperatures imparted by collisions represent a region of parameter space between internally-heated magma oceans, as in this study, and whole-sale atmospheric blowoff.

## 8. Conclusions

We find that $^{26}$Al decay is sufficient to cause short-lived silicate magma oceans on planetesimals of sufficient mass that accreted during the first million years of solar system history. Vapor pressure buildup adjacent to the surfaces of the evaporating magmas would have inevitably led to an approach to near-equilibrium isotope partitioning between the vapor phase and the silicate melt even at pressures well below those thought to have obtained in the protostellar nebula. Numerical simulations of the evaporation process show that a steady-state far-field vapor pressure of $10^{-8}$ bar forms for evaporating bodies of order ½ $M_{\mathrm{Pluto}}$, and that the vapor pressure at the surface of the evaporating body is $10^{-4}$ bar, corresponding to 95% saturation. These results suggest that planetesimals massive enough to retain rock vapor will apporach vapor/melt equilibrium at the surface. For all masses, evaporation into a protostellar gas of order $10^{-4}$ bar results in 99% saturation at the vapor/melt interface.

We modeled the Si and Mg isotopic composition of bulk Earth as a consequence of accretion of planetesimals that evaporated subject to the conditions described above. Our results suggest that the best fit to bulk Earth is for carbonaceous-chondrite like source materials evaporating into the solar protostellar disk, assuming a pressure of $H_2$–dominated gas of about $10^{-4}$ bar, on timescales of $10^4$ years. In the context of this scenario, evaporation led to a 12% loss of Mg and a 15% loss of Si. Under these conditions vapor/melt fractionation is, for all practical purposes, indistinguishible from equilibrium. Enstatite chondrite starting materials do not fit the combination of isotope ratios and mass losses evidenced by the bulk composition of the silicate Earth.

The isotopic effects of rock vapor atmospheric escape may contribute to the heavy isotope enrichment of the molten rock. However, this explanation requires a nearly complete decoupling between mass loss and isotope fractionation.

After evaluating some of the complexities involved, one arrives at the relatively straightforward conclusion that inheritance of the isotopic signatures of near-equilibrium isotope fractionation between vapor and melt should have been the norm during rocky planet formation. These conditions were imposed by steady-state low-pressure rock vapor atmospheres around larger planetesimals or by the presence of protostellar gas. Scenarios involving free evaporation into vacuum do not appear to be viable.

The results of this analysis imply that Earth was composed of ~ 75% evaporated materials and 25% unevaporated materials, and that the budget of moderately volatile elements is dominated by the latter fraction.

**Acknowledgments**: E.D.Y. acknowledges financial support from NASA grant NNX15AH43G issued through the Emerging Worlds program. F. N. acknowledges support from NASA grant NNH17ZDA001N-EW. H.E.S. gratefully acknowledges support from NASA under grant numbers 17-XRP17_2-0055 and 14-XRP14_2-0141 issued through the Exoplanets Research Program. A.S. acknowledges support from the Carnegie Institution for Science.

## Appendix A: Supplmentary Data

**Table A1.** Bulk chemical compositions used for starting materials and bulk silicate Earth.

| Wt % | CI[1] | CI Fe free | EL Avg[2] | EL Fe free | Bulk Earth[3] | Bulk Earth Fe free |
|---|---|---|---|---|---|---|
| **$SiO_2$** | 34.00 | 52.99 | 41.51 | 60.55 | 34.44 | 51.230 |
| **$TiO_2$** | 0.11 | 0.17 | 0.1 | 0.15 | 0.00 | 0.002 |
| **$Al_2O_3$** | 2.50 | 3.90 | 2.14 | 3.12 | 3.00 | 4.469 |
| **$Cr_2O_3$** | 0.00 | 0.00 | 0.27 | 0.39 | 0.00 | 0.010 |
| **$Fe_2O_3$** | 0.00 | 0.00 | 0 | 0.00 | 0.000 | 0.000 |
| **FeO** | 36.16 | 0.00 | 0 | 0.00 | 0.000 | 0.000 |
| **MnO** | 0.39 | 0.61 | 0.15 | 0.21 | 0.00 | 0.002 |
| **MgO** | 24.58 | 38.31 | 22.43 | 32.72 | 25.54 | 37.986 |
| **CaO** | 2.03 | 3.16 | 0.69 | 1.00 | 2.39 | 3.559 |
| **$Na_2O$** | 1.04 | 1.62 | 0.82 | 1.19 | 0.24 | 0.361 |
| **$K_2O$** | 0.10 | 0.16 | 0.09 | 0.13 | 0.02 | 0.000 |
| **$P_2O_5$** | 0.00 | 0.00 | 0.23 | 0.34 | | 0.000 |
| **$H_2O+$** | 0.00 | 0.00 | 0.23 | 0.35 | | 0.000 |
| **$H_2O-$** | 0.00 | 0.00 | 0.14 | 0.20 | | 0.000 |
| **Fe(m)** | | 0.00 | 20.21 | 0.00 | 32.00 | 0.000 |
| **Ni** | | 0.00 | 1.53 | 0.00 | 0.02 | 0.027 |
| **Co** | | 0.00 | 0.08 | 0.00 | | 0.000 |
| **FeS** | 0.00 | 0.00 | 8.57 | 0.00 | | 0.000 |
| **C** | 0.00 | 0.00 | 0.25 | 0.36 | | 0.000 |
| **Other** | 0.00 | 0.00 | 0.93 | 0.00 | | 0.000 |
| **Total** | 100.91 | 100.91 | 100.34 | 100.70 | 97.66 | 97.65 |

1. Newsom (1995); 2. Jarosewich (1990); 3. McDonough and Sun (1995)

**Table A2**. Parameters for four evaporation models as evaluated at $t$ = 50 seconds into the evaporation process.

| | **Model 1: Steady-state atmosphere** | **Model 2: Larger body in protostellar gas** | **Model 3: Smaller body in protostellar gas** | **Model 4: Free evaporation** |
|---|---|---|---|---|
| Melt Radius (km) | 700 | 700 | 50 | 50 |
| $PH_2$ (far field) bar | 1.000E-12 | 1.000E-04 | 1.000E-04 | 1.00E-12 |
| $P$ total (far field) bar | 1.000E-08 | 1.083E-04 | 1.040E-04 | 1.04E-11 |
| T (K) | 2000 | 2000 | 2000 | 2000 |
| d$s$/d$t$ (cm/s) | 6.801E-07 | 5.125E-07 | 5.264E-07 | 1.263E-05 |
| wt%MgO | 38.31 | 38.31 | 38.31 | 38.31 |
| wt%$SiO_2$ | 52.99 | 52.99 | 52.99 | 52.99 |
| wt%CaO | 3.16 | 3.16 | 3.16 | 3.16 |
| w%$Al_2O_3$ | 5.54 | 5.54 | 5.54 | 5.54 |
| | | | | |
| $J_{evap}$ $^{24}$Mg (moles cm$^{-2}$ s$^{-1}$) | 1.099E-07 | 8.317E-06 | 8.208E-06 | 1.099E-07 |
| $J_{net}$ $^{24}$Mg (moles cm$^{-2}$ s$^{-1}$) | 6.091E-09 | 4.483E-09 | 4.553E-09 | 1.084E-07 |
| PMg saturation (bar) | 1.833E-05 | 1.388E-03 | 1.370E-03 | 1.833E-05 |
| PMg (bar) adjacent surface | 1.732E-05 | 1.387E-03 | 1.369E-03 | 2.502E-07 |
| Saturation | 0.94457 | 0.99946 | 0.99945 | 0.01365 |
| | | | | |
| $J_{evap}$ $^{28}$Si (moles cm$^{-2}$ s$^{-1}$) | 5.226E-07 | 3.956E-05 | 3.956E-05 | 5.226E-07 |
| $J_{net}$ $^{28}$Si (moles cm$^{-2}$ s$^{-1}$) | 2.769E-08 | 2.070E-08 | 2.124E-08 | 5.124E-07 |
| $P$SiO saturation (bar) | 1.181E-04 | 8.940E-03 | 8.938E-03 | 1.181E-04 |
| $P$SiO (bar) adjacent surface | 1.118E-04 | 8.935E-03 | 8.933E-03 | 2.302E-06 |
| Saturation | 0.94701 | 0.99948 | 0.99946 | 0.01950 |
| | | | | |
| $J$ Ca (moles cm$^{-2}$ s$^{-1}$) | 2.761E-12 | 2.386E-12 | 7.420E-10 | 1.781E-08 |
| $J$ Al total (moles cm$^{-2}$ s$^{-1}$) | 1.761E-20 | 1.960E-24 | 1.431E-09 | 3.435E-08 |
| Total gas flux (moles cm$^{-2}$ s$^{-1}$) | 2.909E-08 | 2.518E-08 | 2.587E-08 | 6.208E-07 |
| | | | | |
| $J$ $^{24}$Mg/ $J$ $^{24}$Mg$_{congruent}$ | 0.37272 | 0.36803 | 0.36877 | 0.36083 |
| $J$ $^{28}$Si/ $J$ $^{28}$Si$_{congruent}$ | 1.82470 | 1.82970 | 1.82890 | 1.83750 |
| $J$ Ca/ J Ca$_{congruent}$ | 0.00330 | 0.00330 | 0.00330 | 0.00344 |
| $J$ Al$_{total}$/$J$ Al$_{total,\ congruent}$ | 1.092E-11 | 1.407E-15 | 2.050E-14 | 2.295E-07 |
| | | | | |
| $\alpha_{eq}$ (melt/vapor) $^{25}$Mg/$^{24}$Mg | 1.000274 | 1.000274 | 1.000274 | 1.000274 |
| $\alpha_{evap}$ (vapor/melt) $^{25}$Mg/$^{24}$Mg | 0.986900 | 0.986900 | 0.986900 | 0.986900 |
| $\alpha_{net}$ (vapor/melt) $^{25}$Mg/$^{24}$Mg | 0.999106 | 0.999719 | 0.999719 | 0.987073 |
| | | | | |
| $\alpha_{eq}$ (melt/vapor) $^{29}$Si/$^{28}$Si | 1.000617 | 1.000617 | 1.000617 | 1.000617 |
| $\alpha_{evap}$ (vapor/melt) $^{29}$Si/$^{28}$Si | 0.989800 | 0.989800 | 0.989800 | 0.989800 |
| $\alpha_{net}$ (vapor/melt) $^{29}$Si/$^{28}$Si | 0.998871 | 0.999379 | 0.999378 | 0.989985 |
| D $^{24}$Mg gas (cm$^2$ s$^{-1}$) | 3.6015E+08 | 34585.07 | 36015.02 | 3.6015E+11 |
| D $^{28}$Si gas (cm$^2$ s$^{-1}$) | 1.8504E+08 | 17771.53 | 18506.30 | 1.8506E+11 |